\documentclass[a4paper,onecolumn,oneside,11pt]{article}
\usepackage{geometry}
\usepackage{amsmath, amsfonts,amssymb,theorem,
euscript,array,enumerate,amsfonts,mathrsfs,color}
\usepackage{graphics}
\usepackage{epstopdf}
\usepackage[dvips]{epsfig}
\input xy
\xyoption{all}

\definecolor{Ying}{rgb}{0.8,0,0.5}

\newcommand{\bde}{\begin{displaymath}}
\newcommand{\ede}{\end{displaymath}}
\newcommand{\el}{\end{lem}}
\newcommand{\be}{\begin{equation}}
\newcommand{\ee}{\end{equation}}
\newcommand{\beq}{\begin{eqnarray*}}
\newcommand{\eeq}{\end{eqnarray*}}
\newcommand{\beqa}{\begin{eqnarray}}
\newcommand{\eeqa}{\end{eqnarray}}
\newcommand{\bel }{\left\{\begin{array}{ll}}
\newcommand{\eel}{\cr \end{array} \right.}
\newcommand{\bex}{\begin{ex} \rm }
\newcommand{\eex}{\end{ex}}

\newcommand{\bp}{\begin{pro}}
\newcommand{\ep}{\end{pro}}

\def\hat{\widehat}

\def\beqlb{\begin{eqnarray}}\def\eeqlb{\end{eqnarray}}
\def\beqnn{\begin{eqnarray*}}\def\eeqnn{\end{eqnarray*}}
\def\mbf{\mathbf}

\theoremstyle{plain}
\theorembodyfont{\upshape\it}
\newtheorem{Thm}{\bf Theorem}[section]
\newtheorem{Pro}[Thm]{\bf Proposition}
\newtheorem{Lem}[Thm]{\bf Lemma}
\newtheorem{Cor}[Thm]{\bf Corollary}
\theorembodyfont{\upshape}

\newcommand {\proof} {\noindent {\sc Proof:  }}
\newcommand {\finproof} {\hfill $\Box$ \vskip 5 pt }

\def\edoc{\end{document} }

\begin{document}
%\baselinestretch{1.1}
\thispagestyle{empty}
\author{Ying Jiao \footnote{Universit\'e Claude Bernard-Lyon 1, Institut de Science Financier et d'Assurances, 50 Avenue Tony Garnier, 69007 Lyon, France and Peking University, BICMR, 100871 Beijing China. \texttt{ying.jiao@univ-lyon1.fr}} \and Chunhua Ma  \footnote{Nankai University, School of Mathematical Sciences and LPMC,  300071 Tianjin, China.  \texttt{mach@nankai.edu.cn}} \and Simone Scotti \footnote{Universit\'e Paris Diderot-Paris 7, Laboratoire de Probabilit\'es et Mod\`eles Al\'eatoires, Site Sophie Germain,  75013 Paris, France.  \texttt{scotti@math.univ-paris-diderot.fr}}\and Chao Zhou \footnote{National University of Singapour. \texttt{matzc@nus.edu.sg}}}

\title{The Alpha-Heston Stochastic Volatility Model}
\date{\today}
\maketitle

\begin{abstract}
We introduce an affine extension of the Heston model where the instantaneous variance process contains a jump part driven by $\alpha$-stable processes with $\alpha\in(1,2]$. In this framework, we examine the implied volatility and its asymptotic behaviors for both asset and variance options. Furthermore, we examine the jump clustering phenomenon observed on the variance market  and provide a jump cluster decomposition  which allows to analyse the cluster processes.    

\vspace{3mm}

{\bf MSC:} 91G99, 60G51, 60J85 \\
\vspace{-4mm}

{\bf Key words:} Stochastic volatility and variance, affine models, CBI processes,  implied volatility surface, jump clustering. 
\end{abstract} 

%\documentclass[11pt,a4paper]{article}
%\usepackage{a4wide,amsfonts,amsmath,latexsym,amsthm,amssymb,euscript,eufrak,graphicx,units,mathrsfs, color,setspace}

%\usepackage[active]{/usr/share/doc/HTML/en/kdvi/srcltx}
%\usepackage[dvips,final]{graphicx}
%\pagestyle{headings}

%\usepackage{graphicx}

%\usepackage{subfigure}
%\usepackage[french,english]{babel}
%\usepackage[T1]{fontenc}

%\DeclareMathAlphabet{\eufrak}{U}{}{}{}  % Euler fraktur math
%\SetMathAlphabet\eufrak{normal}{U}{euf}{m}{n}
%\SetMathAlphabet\eufrak{bold}{U}{euf}{b}{n}
%\newcommand{\ndot}{\raisebox{.4ex}{.}}

\section{Introduction}

\label{intro}

The stochastic volatility models have been widely studied  in literature and one important approach consists of the Heston model \cite{Hes1993} and its extensions. In the standard Heston model, the instantaneous variance is a square-root mean-reverting CIR (Cox-Ingersoll-Ross \cite{CIR85}) process. On one hand, compared to the Black-Scholes framework, Heston model has the advantage to  reproduce some stylized facts in equity and foreign exchange option markets. The model provides analytical tractability of pricing formulas which allows for efficient calibrations. On the other hand, the limitation of  Heston model has also been carefully examined. For example, it is unable to produce extreme paths of volatility during the crisis periods, even with very high volatility of volatility (vol-vol) parameter. In addition, the Feller condition, which is assumed in Heston model to ensure that the volatility remains strictly positive, is often violated in practice, see e.g. Da Fonseca and Grasselli \cite{DaFGra11}.    

To provide more consistent results with empirical studies, a natural extension is to consider jumps in the stochastic volatility models.  In the Heston framework, Bates \cite{Bates1996} adds jumps in the dynamics of the asset, while Sepp \cite{Sepp2008} includes jumps in both asset returns and the variance, both papers using Poisson processes. In Barndorff-Nielsen and Shephard \cite{BNS2001}, the volatility process is the superposition of a family of positive non-Gaussian Ornstein-Uhlenbeck processes.  Nicolato et al. \cite{NPS2017} study the case where a jump term is added to the instantaneous variance process which depends on an increasing and driftless L\'evy process,  and they analyze the impact of jump diffusions on the realized variance smile and the implied volatility of VIX options.  More generally, Duffie et al. \cite{DPS2000} \cite{DFS2003} propose the affine jump-diffusion  framework for the asset and stochastic variance processes. %The moment explosion  and long-term behaviors of the affine  stochastic volatility models have been  studied by Keller-Ressel \cite{KR11}.  
There are also other extensions of Heston model. Grasselli \cite{G2016} combines standard Heston model with the so-called $3/2$ model where the volatility is the inverse of the Heston one. Kallsen et al \cite{KMV2011} consider the case where stock evolution includes a time-change L\'evy process.  In the framework of rough volatility models (see for example El Euch et al. \cite{EFR2016} and  Gatheral et al. \cite{GJR2014}), El Euch and Rosenbaum \cite{ER2016} propose the rough Heston model where the Brownian term is replaced by a fractional Brownian motion and they provide the characteristic function by using the fractional Riccati equation.

\begin{figure}[h]
\caption{The CBOE's VIX value from January 2004 to July 2017.}
\label{fig:vix}
\begin{center}
\includegraphics[width=0.65\textwidth]{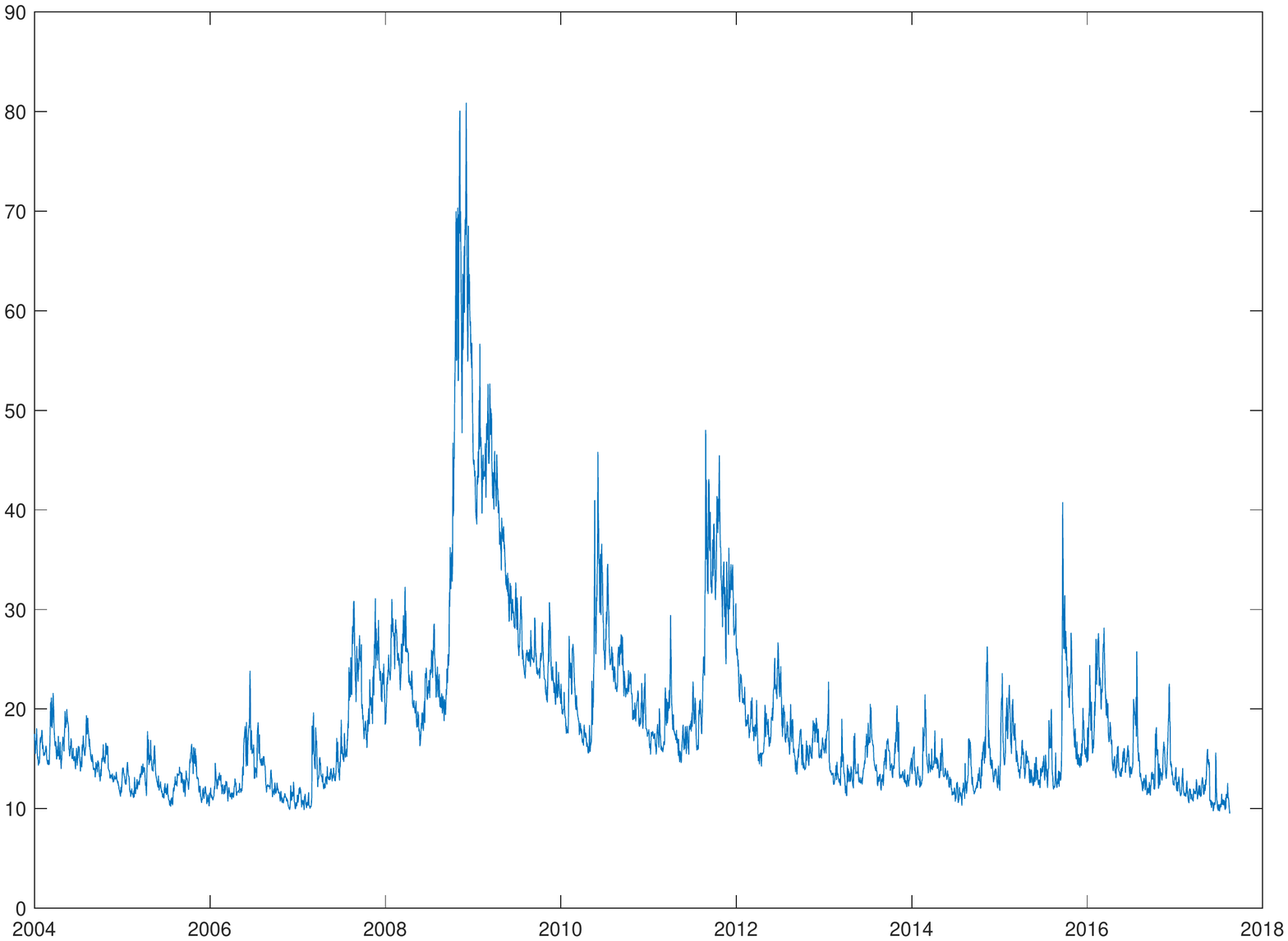}
\end{center}
\end{figure}

In this paper, we introduce  an extension of Heston model, called the $\alpha$-Heston model, by adding a self-exciting jump structure in the instantaneous variance. 
On financial markets, %the  volatility data often exhibit the phenomenon of jump clusters. T
the CBOE's Volatility Index (VIX) has been introduced as a measure of market volatility of S\&P500 index. Starting from 2004, this index is exchanged via the VIX futures,  %where the derivatives of the realized variance and volatility, including options on the VIX index, 
and its derivatives have  been developed quickly in the last decade. Figure \ref{fig:vix} presents the daily closure values of VIX index from January 2004 to July 2017. 
 The historical data shows clearly that the VIX can have very large variations and  jumps, particularly during the periods of crisis and partially due to the lack of ``storage''.  Moreover the jumps occur frequently in clusters. We note several significant jump clusters, the first one associated to the subprime crisis during 2008-2010, the second associated to the sovereign crisis of Greece during 2010-2012, and the last one to the Brexit event around 2016-2017.  Between the jump clusters, the VIX values drop to  relatively low levels during normal periods. One way to model the cluster effect in finance is to adopt the Hawkes processes \cite{Hawkes} where it needs to specify the jump process together with its intensity. So the inconvenience is that the dimension of the  concerned stochastic processes is increased. For the volatility data, El Euch et al. \cite{EFR2016} emphasize that the market is highly endogenous and justify the use of nearly unstable Hawkes processes in their framework. Furthermore,  Jaisson and Rosenbaum \cite{JR15} prove that nearly unstable Hawkes processes converge to a CIR process  after suitable rescaling. Therefore it is natural to reconcile the Heston framework with a suitable jump structure in order to describe the jump clusters. %and at the same time preserve the main advantages of the Heston type models. 

Compared to the standard Heston model, the $\alpha$-Heston model includes  an $\alpha$-root term and  a compensated $\alpha$-stable L\'evy process in the stochastic differential equations (SDE) of the instantaneous variance process $V= (V_t,t\geq 0)$.  The number of extra parameters  is sparing and the only main parameter $\alpha$ determines the jump behavior. This model allows to describe the cluster effect in a parsimonious and coherent way. We adopt a related approach  of continuous-state branching processes with immigration (CBI processes). With the general integral characterization for SDE in Dawson and Li \cite{DawsonLi}, $V$ can be seen as a marked Hawkes process with infinite activity influenced by a Brownian noise (see Jiao et al. \cite{JMS2017}), which is suitable to model the self-exciting jump property.  In this model,  the $\alpha$-stable jump process is leptokurtotic and heavy-tailed. The parameter $\alpha$ corresponds to the Blumenthal-Getoor index. Hence it's able to seize both large and small fluctuations  and even extreme high peaks during the crisis period. In addition, the law of jumps follows the Pareto distribution. Empirical regularities in economics and finance often suggest the form of Pareto law: Liu et al. \cite{Liu99} found that the realized volatility matches with a power law tail;  more recently, Avellaneda and Papanicolaou \cite{AP2018} showed that the right-tail distribution of VIX time series can be fitted to a Pareto law.  We note that the same Feller condition applies as in the standard Heston case and this condition is more easily respected by the $\alpha$-Heston model since the behavior of small jumps with infinite activity is similar to a Brownian motion so that the  jump part  allows to reduce the vol-vol parameter.

%The $\alpha$-Heston model aims to preserve  main advantages of Heston model and to overcome some of its empirical inconsistency.  First, our model has an explicit formula for the characteristic function of the asset log-price, the variance and its integrated process. It is explicit up to a numerical integral. Thank to this formula, fast model calibration and computation of contingent claim prices is allowed by efficient numerical methods in the spirit of Carr and Madan \cite{CM1999}. We want to preserve the usual parameters of Heston model, since they have a financial explanation in term of principal component analysis. Indeed the initial volatility arranges the whole volatility curve. The correlation is linked to the skew and to the so-called leverage effect and the volatility of volatility (vol-vol) parameter deals with the smile.    We use the transform analysis of affine jump models to obtain analytic results for the characteristic function of the asset log-price, which is explicit up to a numerical integral. 

Thanks to the link between CBI and affine processes established by Filipovi\'c \cite{F01}, our model belongs to the class of  affine jump-diffusion models in  Duffie et al. \cite{DPS2000}, \cite{DFS2003} and the general result on the characteristic functions holds for the $\alpha$-Heston jump structure. %Although the general link between CBI processes and affine models have been established by Filipovi\'c \cite{F01}, the special $\alpha$-Heston jump structure has drew little attention in previous affine stochastic volatility literature  probably because on one hand, the explicit SDE related to $\alpha$-Heston model has been introduced more recently by Fu and Li \cite{FL10} and Li and Ma \cite{LM13} and on the other hand, the associated generalized Riccati operator is not analytic, which breaks down some arguments borrowed from complex analysis to describe the model.
However, the associated generalized Riccati operator is not analytic, which breaks down certain arguments borrowed from complex analysis.
%Fortunately, the main result about the Fourier/Laplace transform in \cite{DPS2000} remains valid,  with which we obtain an explicit formula for the characteristic function of the triplet of asset log-price, the variance and its integrated process up to a numerical integral. We can also examine  the moment explosion and asymptotic behaviours based on results in Keller-Ressel \cite{KR11}.
One important point is that although theoretical results on generalized Riccati operators  are established for general affine models, in many explicit examples, the generalized Riccati equation which is associated to the state-dependent variable of $V$ is quadratic. 
%It results in a lack of flexibility of the cumulant generator function of the variance process and then some preclusion to fit financial data such as implied volatility for options. 
The $\alpha$-Heston model allows to add more flexibility to the cumulant generator function since its generalized Riccati operator contains a supplementary $\alpha$-power term. 
We examine  the moment explosion behaviors of both asset and variance processes following Keller-Ressel \cite{KR11}.
We are also interested in the implied volatility surface and its asymptotic behaviors based on  the model-free result of Lee \cite{L04}. For the asset options, we show that the  wing behaviors of the volatility smile at extreme strikes are the sharpest. For the variance options, we first estimate the asymptotic property of tail probability of the variance process. 
%I COMMENT UNTILL WE WILL OBTAIN BETTER GRAPHS  
%Then by examining the different behaviors of left and right wings respectively, we see that  the volatility surface at extreme strikes of variance options is characterized by an upward-sloping smile, as suggested by literature.

One of the most interesting features of the $\alpha$-Heston model is that by using the CBI characteristics as in Li and Ma \cite{LM13}, we can thoroughly analyse the jump cluster effect. %, that is, the phenomenon that a large jump may induce a cluster of consecutive jumps. 
Inspired by Duquesne and Labbe \cite{DuqLab}, we provide a decomposition formula for the variance process $V$ which contains a fundamental part together with a sequence of jump cluster processes. This decomposition implies a branching structure in the sense that each cluster process is  induced by a ``mother jump'' which is followed by ``child jumps''. The mother jump represents  a triggering shock on the market and is driven by exogenous  news in general whereas the child jumps may reflect certain contagious effect. We then study relevant properties such as the duration of one cluster and the number of clusters occurred in a given period. We are particularly interested in the role played by the main parameter $\alpha$. %To illustrate its impact, we provide some numerical simulations. 

The rest of the paper is organized as follows.  We present the model framework in Section \ref{sec:Model}. Section \ref{sec: affine} is devoted to the affine characterization of the model and related properties. In Section \ref{sec: extreme strike}, we study the asymptotic implied volatility behavior of asset and variance options. Section \ref{sec: clusters} deals with the analysis of jump clusters. We conclude the paper by providing the proofs in Appendix.  

\section{Model framework}
\label{sec:Model}

Let us fix a probability space $(\Omega, \mathcal A, \mathbb Q)$ equipped with a filtration $\mathbb F=(\mathcal F_t)_{t\geq 0}$ which satisfies the usual conditions. We first present a family of stochastic volatility models by using a general integral representation of SDEs with random fields. %, see for instance \cite{KM90}.  
Consider  the asset price process $S=(S_t,t\geq 0)$ given by  
\begin{equation}\label{Heston-integral}
\frac{d S_t}{S_t} =   rdt + \int_0^{V_t} B(dt,du), \quad S_0>0 \end{equation}
where $r\in\mathbb R_+$ is the constant interest rate, $B(ds,du)$ is a white noise on $\mathbb{R}_+^2$ with intensity $dsdu$, and the process $V=(V_t,t\geq 0)$ is given  by 
\begin{equation}\label{Vol integral}V_t =  V_0 + \int_0^t a( b  - V_s  ) ds + \sigma \int_0^t \int_0^{V_s} W(ds,du) + \sigma_N \int_0^t \int_0^{V_{s-} } \int_{\mathbb{R}^+} \zeta \widetilde{N}(ds,du, d\zeta) 
\end{equation}
where $a,b,\sigma, \sigma_N\in\mathbb R_+$, $W(ds,du)$ is a white noise on $\mathbb{R}_+^2$ correlated  to $B(ds,du)$ such that $B(ds,du) = \rho W(ds,du) +\sqrt{1-\rho^2} \overline{W}(ds, du)$ with $\overline{W}(ds,du)$ being an independent white noise and $\rho\in (-1,1)$, $\widetilde N(ds,du,d\zeta)$ is an independent compensated Poisson random measure on $\mathbb{R}_+^3$ with intensity $dsdu\nu(d\zeta)$ with $\nu(d\zeta)$ being a L\'evy measure on $\mathbb{R}_+$ and satisfying  
$\int_0^\infty (\zeta\wedge\zeta^2)\nu(d\zeta)<\infty $. The measure $\mathbb Q$ stands for the risk-neutral probability measure. We shall discuss in more detail the change of probability in Section \ref{sec: change of probability}. 

The variance process $V$ defined above is a CBI process (c.f. Dawson and Li \cite[Theorem 3.1]{DawsonLi}) with  the branching mechanism given by
\begin{equation}\label{equ: Psi general}
\Psi(q)=aq +\frac{1}{2}\sigma^2q^2+ \int_0^{\infty}(e^{-q\sigma_N\zeta}-1+q\sigma_N\zeta){\nu}(d\zeta)
\end{equation}
and the immigration rate $\Phi(q)=a b q$. The existence and uniqueness of a strong solution of  \eqref{Vol integral} is proved in  \cite{DawsonLi} and \cite{LM13}. From  the financial viewpoint, Filipovi\'c \cite{F01} has shown how the CBI processes naturally enter the field of affine term structure modelling. The integral representation provides a  family of processes where the integral intervals in \eqref{Vol integral} depend on the value of the process itself, which means that the jump frequency will increase when a jump occurs, corresponding to the self-exciting property. 
 
We are particularly interested in the following model, which is called  the {$\alpha$-Heston model},
\begin{eqnarray}\label{alpha Heston-root}
\displaystyle \frac{d S_t}{S_t} &=& \displaystyle   r dt + \sqrt{V_t} dB_t \\
dV_t &=& a\left ( b - V_t  \right) dt + \sigma  \sqrt{V_t} dW_t
+\sigma_N \sqrt[\alpha]{V_{t-}}  dZ_t \label{alpha CIR} 
\end{eqnarray}
where $B=(B_t,t\geq 0)$  and $W=(W_t,t\geq 0)$ are correlated Brownian motions $d\left<B,W\right>_t=\rho dt$ and $Z=(Z_t,t\geq 0)$ is an independent spectrally positive compensated $\alpha$-stable  L\'evy process with parameter $\alpha \in (1,2]$ whose Laplace transform is given, for any $q\geq 0$, by
 \beqnn
 \mathbb{E}\big[e^{-qZ_t}\big]=\exp\Big(-\frac{tq^\alpha}{\cos(\pi\alpha/2)}\Big).
 \eeqnn 
The equation \eqref{alpha CIR} corresponds to the choice of the L\'evy measure   \beqlb\label{Levymeasure} 
\nu_\alpha(d\zeta)=-{1_{\{\zeta>0\}} d\zeta \over \cos(\pi\alpha/2)\Gamma(-\alpha)\zeta^{1+\alpha}}, \quad 1<\alpha<2.
 \eeqlb
in \eqref{Vol integral}. Then the solutions of the two systems of SDEs  admit the same probability law and are equal almost surely in an expanded probability space by \cite{Li11}. 

The {$\alpha$-Heston model} is an extension of standard Heston model in which  the jump part of the variance process depends on an $\alpha$-square root jump process.  In particular, we call the process $V$ defined in \eqref{alpha CIR} an $\alpha$-{CIR}$(a,b,\sigma,\sigma_N,\alpha)$ process and the existence and uniqueness of the strong solution are established in Fu and Li \cite{FL10}. In this case, by \eqref{equ: Psi general} and \eqref{Levymeasure}, the variance $V$ has  the explicit branching mechanism   
 \begin{equation}\label{equ: Psi SCIR}
 \Psi_{\alpha}(q)=a q+\frac{\sigma^2}{2}q^2-\frac{\sigma_N^\alpha}{\cos(\pi\alpha/2)}q^\alpha.
 \end{equation} Compared to the standard Heston model, the parameter $\alpha$ characterizes the jump behavior and the tail fatness of the instantaneous variance process $V$. When $\alpha$ is near $1$, $V$ is more likely to have large jumps but its values between large jumps tend to be small due to deeper negative compensations (c.f. \cite{JMS2017}).   When $\alpha$ is approaching $2$, there will be less large jumps but more frequent small jumps.  In the case when $\alpha=2$, the process $Z$ reduces to an independent Brownian motion scaled by $\sqrt{2}$ and the model is reduced to a standard Heston one. 
 
The Feller condition, that is, the inequality $2ab\geq\sigma^2$, is often assumed in the Heston model to ensure the positivity of the process $V$. In the $\alpha$-Heston model, the same condition remains to be valid. More precisely, for any $\alpha\in(1,2)$, the point $0$ is an inaccessible boundary for \eqref{alpha Heston-root} if and only if $2ab\geq\sigma^2$ for any $\sigma_N\geq 0$ (c.f. \cite[Proposition 3.4]{JMS2017}).  From the financial point of view, this means that the jumps have no impact on the possibility for the volatility to reach the origin, which can be explained by the fact that only positive jumps are added and their compensators are proportional to the process itself. When $\alpha=2$, the Feller condition becomes $2ab\geq\sigma^2+2\sigma_N^2$ since $Z$ becomes a scaled Brownian motion. Empirical studies  show that (see e.g. Da Fonseca and Grasselli \cite{DaFGra11}, Graselli \cite{G2016}), in practice, the Feller condition is often violated since when performing  calibrations on equity market data high  vol-vol  is required to reproduce large variations. This point is often seen as a drawback of the Heston model. In the $\alpha$-Heston model,  part of the vol-vol parameter  is seized  
by the jump part. Indeed, as shown by Asmussen and Rosinski \cite{AR2001},  the small jumps of a L\'evy process can be approximated by a Brownian motion, so that  the small jumps induced by the infinite activity of the variance process generates a behaviour similar as that of a Brownian motion. This allows to reduce mechanically the contribution from the Brownian part and hence the vol-vol parameter. As a consequence, our model is more likely to  preserve the Feller condition and the positivity of the volatility process. 
 
\begin{figure}
\caption{Simulation of the variance process $V$.}
\begin{center}
\includegraphics[width=0.6\textwidth]{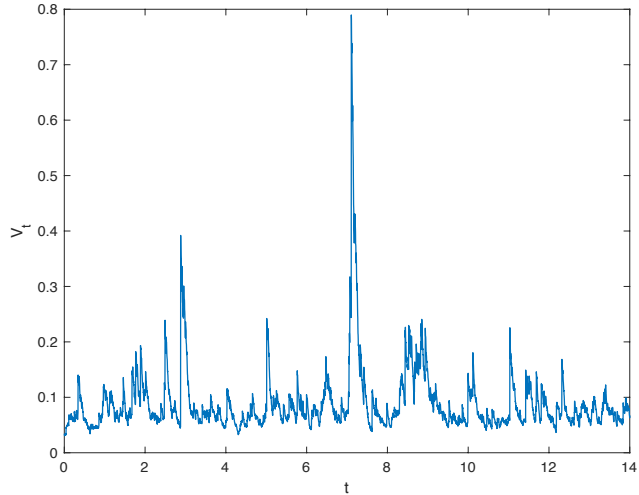}
\end{center}
\label{fig: variance}       
\end{figure}

Figure \ref{fig: variance}   provides a simulation of the variance process $V$ defined in \eqref{alpha CIR} for a period of $T=14$, in comparison with the empirical VIX data (from 2004 to 2017) in Figure \ref{fig:vix}. The parameters are chosen to be $a=5$, $b=0.14$, $\sigma=0.08$, $\sigma_Z=1$ and $\alpha=1.26$. The initial value is fixed to be $V_0=0.03$ according to the VIX data on January 2nd, 2004. Note that the Feller condition is largely satisfied with the above choice of parameters and  the values of $V$  are always positive in Figure \ref{fig: variance}. We also observe the cluster phenomenon for jumps and in particular some large jumps concentrated on a short period. At the same time, the values of the variance process $V$ remain to be at a relatively low level between the jumps, which corresponds to the normal periods between the crisis, similarly as shown by empirical data in Figure \ref{fig:vix}.

\section{Affine characteristics}\label{sec: affine}
%The $\alpha$-Heston model belongs to the affine jump-diffusion models. 
In this section, we give the joint Laplace transform of the log-price, the variance  and its integrated process according to Duffie et al. \cite{DPS2000, {DFS2003}} and Keller-Ressel \cite{KR11}.
We begin by discussing the probability change between the historical and the risk-neutral pricing probability measures. We shall also  make comparisons with several other affine models in literature. 

\subsection{Change of probability measures}\label{sec: change of probability}

We have assumed that model dynamics   \eqref{Heston-integral}, \eqref{Vol integral} and (\ref{alpha Heston-root}) are specified under a risk-neutral probability $\mathbb Q$. However, it is important  to establish a link with the physical or historical one generally denoted by $\mathbb{P}$ in order to keep a tractable form for the evolution of the processes describing the market.  
%Our framework shares the main drawback of jump diffusion model, that is the model is incomplete. However, this drawback is shared with the recent models. Considering for instance rough volatility model the whole forward variance curve, that is an infinite and uncountable number of contingent claims, is needed to complete the market, see section 3.2 in \cite{ER2016}.
The construction of  an equivalent historical probability is based on an Esscher-type transformation in Kallsen et al. \cite{KMK2010} which is a natural extension of the class proposed by Heston \cite{Hes1993}. %Moreover, it shares the same behaviour since the main change concerns the parameter on mean reverting speed. 
The next result shows that the general class of temperated Heston-type model  is closed under 
the change of probability and is a slight modification of \cite[Proposition 4.1]{JMS2017}. 
 
\begin{Pro}\label{pro:changementprob} Let $(S, V)$ be as in \eqref{Heston-integral} and \eqref{Vol integral} under the probability measure $\mathbb Q$ 
and assume that the  filtration $\mathbb F$ is generated by the random fields $(W, \overline{W})$ and $\widetilde N$.  Fix 
$(\eta, \overline{\eta})\in\mathbb{R}^2$ and $\theta\in\mathbb{R}_+$, and define
 \beqnn
 U_t:=\eta\int_0^t\int_0^{V_s}W(ds,du) + \overline{\eta}\int_0^t\int_0^{V_s}\overline{W}(ds,du) 
 +\int_0^t\int_0^{V_{s-}}\int_0^\infty (e^{-\theta\zeta}-1)\widetilde{N}(ds,du,d\zeta).
 \eeqnn
Then the Dol\'eans-Dade exponential $\mathcal{E} (U)$ is a martingale and the probability 
measure $\mathbb P$ defined by 
 \beqnn
\left. \frac{d\mathbb P}{d\mathbb Q}\right|_{\mathcal{F}_t}=\mathcal{E} (U)_t,
 \eeqnn
 is equivalent to $\mathbb Q$.  
 Moreover, under $\mathbb P$, $(S, V)$ satisfy  \eqref{Heston-integral} and \eqref{Vol integral}  with the parameters $\sigma^{\mathbb P}=\sigma$, $\sigma^{\mathbb P}_N=\sigma_N$,
$$a^{\mathbb P}=a-\sigma\eta -\frac{\alpha\sigma_N}{\cos(\pi\alpha/2)}\theta^{\alpha-1}, \quad b^{\mathbb P}=ab/a^{\mathbb P},$$ and the L\'evy measure 
\beqnn 
\nu_\alpha^{\mathbb P}(d\zeta)=-\frac{1_{\{\zeta>0\}}e^{-\theta\zeta}}{\cos(\pi\alpha/2)\Gamma(-\alpha)\zeta^{1+\alpha}}d\zeta.
\eeqnn
\end{Pro}

The model under $\mathbb P$ remains in the CBI class of $\alpha $-Heston model and shares  similar behaviors. 
Note that  the parameters $\eta, \overline{\eta}$ and $\theta$ are chosen such that $a^{\mathbb P}\in\mathbb R_+$. 
As a direct consequence of the above proposition, the return rate of the price process under $\mathbb P$ becomes 
\[\mu^{\mathbb P}_t = r - V_t \left( \rho \eta + \sqrt{1-\rho^2} \overline{\eta} \right).\] 
The risk premiums are given by
\begin{eqnarray*}
\lambda_S(t) &:=& \mu_t^{\mathbb P} -r =  -\left( \rho \eta + \sqrt{1-\rho^2} \overline{\eta} \right) V_t \\
\lambda_V(t) &:=& (a^{\mathbb P}-a) V_t = - \left(\sigma\eta + \frac{\alpha\sigma_N}{\cos(\pi\alpha/2)}\theta^{\alpha-1}\right) V_t   \, .
\end{eqnarray*}
%\end{Cor}
{When $\eta<0$, the risk premium $\lambda_V$ is positively correlated with the volatility process $V$. 
The positive correlation between the risk premium and the volatility can explain the strongly upward sloping in VIX smile detailed in 
\cite{BFG2016}.}

\subsection{Joint characteristic function}

In the Heston model, it is well known  that the characteristic function plays a crucial role for the pricing of derivatives and the model  calibration.  We now provide the joint Laplace transform of the triplet:  the log-price, the variance and its integrated process.  The following result is a direct consequence of \cite{DPS2000} and  \cite{KR11} and its proof is postponed to Appendix.

\begin{Pro}\label{Pro: joint laplace transform}
Let $Y_t=\log S_t$. 
For any $\xi=(\xi_1,\xi_2,\xi_3)\in i\mathbb{R}\times\mathbb{C}_-^2$, 
\begin{equation}
\mathbb E\Big[\exp\big(\xi_1Y_t+\xi_2V_t+\xi_3\int_0^tV_sds\big)\Big]=\exp\Big(\xi_1Y_0+\psi(t,\xi)V_0+\phi(t,\xi)\Big)\end{equation}
where $\phi$ and $\psi$ solve the generalized Riccati equations
\begin{eqnarray}\label{generalized Racci}
\partial_t\phi(t,\xi)&=F(\xi_1,\psi(t,\xi), \xi_3),\quad \phi(0,\xi)=0;\label{generalized Racci 1}\\
\partial_t\psi(t,\xi)&=R(\xi_1,\psi(t,\xi),\xi_3),\quad \psi(0,\xi)=\xi_2.\label{generalized Racci 2}
\end{eqnarray}
Moreover,  the functions $F$ and $R:  i\mathbb{R}\times\mathbb{C}_-^2\rightarrow \mathbb R$ are defined by
\begin{eqnarray}
F(\xi_1,\xi_2,\xi_3)&=&r\xi_1+ab\xi_2,\label{F explicit}\\
\label{R explicit}
R(\xi_1,\xi_2,\xi_3)%&=&\frac{1}{2}q_1^2+\rho\sigma q_1q_2+\frac{1}{2}\sigma^2q_2^2-\frac{1}{2}q_1-aq_2+q_3+\sigma_N^\alpha\int_0^\infty(e^{q_2\zeta}-1-q_2\zeta)\nu_{\alpha}(d\zeta)\\&=&
&=&\frac{1}{2}(\xi_1^2-\xi_1)+\rho\sigma \xi_1\xi_2+\frac{1}{2}\sigma^2\xi_2^2-a\xi_2-\frac{\sigma_N^\alpha}{\cos(\pi\alpha/2)}(-\xi_2)^\alpha+\xi_3.\label{R explicit}
%\ \mbox{Check!}
\end{eqnarray}
\end{Pro}

To compare  the $\alpha$-Heston model with other models in literature, we consider in the remaining of the paper  the usual case as in \cite{DPS2000} and  \cite{KR11} where the third vaiable $\xi_3$ is omitted and $r=0$. Recall that in the standard Heston model, the generalized Riccati operators are given by 
\begin{eqnarray}F_H(\xi_1,\xi_2)=ab\xi_2,\quad { \text{and } }\quad R_H(\xi_1,\xi_2)=\frac{1}{2}(\xi_1^2-\xi_1)+\rho\sigma \xi_1\xi_2+\frac{1}{2}\sigma^2\xi_2^2-a\xi_2.\end{eqnarray}  
By Proposition \ref{Pro: joint laplace transform}, the $\alpha$-Heston model admits
\begin{equation}\label{definition-R}
F(\xi_1,\xi_2)=F_H(\xi_1,\xi_2),\quad { \text{and } }\quad R(\xi_1,\xi_2)=R_H(\xi_1,\xi_2)-\frac{\sigma_N^\alpha}{\cos(\pi\alpha/2)}(-\xi_2)^\alpha.
\end{equation}
Note that the function $R$ in \eqref{definition-R} is not analytic and is well defined only for $\xi_2\leq 0$. The difference $R(\xi_1,\xi_2) -R_H(\xi_1,\xi_2) $ is positive since $\cos(\pi \alpha /2)<0$ for $\alpha\in(1,2]$.  
As stated in  \cite{KR11}, $F$ characterizes the state-independent dynamic of  $(S,V)$ while R characterizes the state-dependent dynamic. In order to  highlight the primacy of function $\psi$ in \eqref{generalized Racci 2}, we refer $R$ as the main generalized Riccati operator. 
%For Heston model, we have that the Riccati operator is analytic, the maximal definition domain of $w_H(\xi_1)$ is given by an ellipse curve which is larger than $[0,1]$ and the explosion time is dichotomized. The explicit formulae can be found in \cite[section 6.1]{KR11}. We now discuss several other models such as  those in Bates \cite{Bates1996}, Barndorff-Nielsen and Shephard \cite{BNS2001} (hereafter BNS) and Nicolato et al. \cite{NPS2017}.

The main point we highlight is that many models discussed in literature admit similar forms of  $R$. In Barndorff-Nielsen and Shephard \cite{BNS2001}, $R$ is a particular case of Heston one,  i.e. 
$\sigma=0$, and the main innovation of their model is to
extend in an interesting way the auxiliary operator $F$. The model in Bates \cite{Bates1996} has a more general generalized Riccati operator $R$ but the new term depends only on the Laplace coefficient
of the stock $S$. So the variance process in \cite{Bates1996} follows the CIR diffusion and hence there is no difference for volatility and variance options compared to Heston model. 
For the stochastic volatility jump model in Nicolato et al. \cite{NPS2017},  the examples  share the same Riccati operator of the Heston model. As  a consequence, the Laplace transform of the variance process has a certain form for the affine function. Then, it is not surprising that ``the specific choice of jump distribution has a minor effect on the qualitative behavior of the skew and the term structure of the implied volatility surface'' as noted in \cite{NPS2017} (see also \cite{NV03}), since the plasticity of the model is limited to the form of the 
auxiliary function $\phi(t,\xi)$ which is independent of the level of initial variance $V_0$ in the cumulant generating function. 

Our model exhibits a different behavior due to the supplementary $\alpha$-power term appearing in the main generalized Riccati operator $R$, which adds more flexibility to the coefficient of the variance $\psi(t,\xi)$ in the cumulant generating function. The reason lies in the fact that the new jump part depends on the variance itself, resulting in a   non-linear dependence in \eqref{R explicit}. 
In other words, the self-exciting property of jump term introduces a completely  different shape of cumulant generating function.

%In summary, we remark that the limit shape of the coefficient of the variance $\psi(t,\xi)$ in the cumulant generating function 
%is mainly the same across a large class of exponential affine models previously detailed in the literature. In other words, the plasticity of the model is limited to the from of the 
%auxiliary function $\phi(t,\xi)$ which is independent of the level of initial variance $V_0$ in the cumulant generating function.  

%{\color{red}
%\begin{remark}[Comparaison with other models]

%\begin{description}
%\item[BNS model] $w_{BNS}(u)$ is well defined for all $u$ and coincides with $w_H(u)$ assuming $\sigma=0$. 

%\item[Bates model] $w_B(u)$ has similar form of  $w_H(u)$ with a topical determinant. 

%\end{description}

%\end{remark}

\section{Asymptotic behaviors and implied volatility}\label{sec: extreme strike}
In this section, we focus on the implied volatility surfaces for both asset and variance options, in particular, on their asymptotic behaviors at small or large strikes. We follow the model-free result in the pioneering paper  of Lee \cite{L04} and aim to obtain some refinements for the specific $\alpha$-Heston model. We also provide the moment explosion conditions.

\subsection{Asset options}

We begin by  providing the following results on the generalized Riccati operator $R$ by \cite{KR11} and give the moment explosion condition for the asset price $S$.
\begin{Pro}\label{proposition-KR}
We assume $a> \sigma \rho$. Define  $w(\xi_1)$  such that $R(\xi_1,w(\xi_1))=0$ and $T_*(u):= \sup\{T:\, \mathbb{E}[S_T^u]<\infty\}$
\begin{enumerate}[(1)]
\item $w(\xi_1)$ has $[0,1]$ as maximal support. 
\item $\forall \xi_1\in[0,1]$ we have $\lim_{t\rightarrow \infty} \phi(t,\xi_1,w) = w(\xi_1)$.  
\item $\forall \xi_1\in[0,1]$ we have $T_*(\xi_1) =\infty$ and $\forall \xi_1\notin[0,1]$ we have $T_*(\xi_1) =0$.
\end{enumerate}
\end{Pro} 
\proof %Recall that $Y_t=\log S_t$ and we have by  (\ref{alpha Heston-root}) that \beqnn dY_t=(r-\frac{1}{2}V_t)dt+\sqrt{V_t}dB_t. \eeqnn
The couple $(Y_t, V_t)$ is an  affine process characterized by (\ref{definition-R}) and $F(u,w) := ab w$. Note that $F(0,0)=R(0,0)=R(1,0)=0$ and 
$\chi(q_1):=\frac{\partial R(q_1,q_2)}{\partial q_2}\big\vert_{q_2=0}=\rho\sigma q_1-a<\infty$. Then by Keller-Ressel \cite[Corollary 2.7]{KR11} we have 
$\mathbb E[S_T]<\infty$ for any $T>0$. Also note that $\chi(0)<0$ and $\chi(1)<0$ as $a>0$, $\rho<0$ and $\sigma>0$. It follows from \cite[Lemma 3.2]{KR11} that
there exist a maximal interval $I$ and a unique function $w\in C(I)\cap C^1(I^\circ)$ such that $R(q_1,w(q_1))=0$ for all $q_1\in I$ with $w(0)=w(1)=0$. Since $0=\sup\{q_2\geq0: R(q_1,q_2)<\infty\}$,
$R(q_1,q_2)>0$ if $q_1<0$ and $q_2<0$, and $R(q_1,0)=\frac{1}{2}q_1(q_1-1)$, we immediately have that $I=[0,1]$. Then the set $\{q_1\in I: F(q_1,w(q_1))<\infty\}$ coincide with $[0,1]$. 
 By \cite[Theorem 3.2]{KR11} we have $\mathbb E[S_T^q]=\infty$ for any $q\in\mathbb{R}\setminus[0,1]$.
\finproof

\begin{Cor}The above proposition implies that for any $T>0$, we have 
\beqnn
\sup\{p>0: \mathbb E[S_T^p]<\infty\}=1\quad  \mbox{and}\quad \sup\{p>0: \mathbb E[S_T^{-p}]<\infty\}=0.
\eeqnn
In other words, the maximal domain of moment generating function $\mathbb E[e^{q\log S_T}]$ is $[0,1]$.
\end{Cor}

Let $\Sigma_S(T, k)$ be the implied volatility of a call option written on the asset price $S$ with maturity $T$ and strike $K = e^k$. Then combined with a model-free result of Lee \cite{L04}, known as the moment formula, it yields that the asymptotic behavior of the implied volatility at extreme strikes is given by 
\begin{equation}\label{Lee formula}
\limsup_{k \rightarrow \pm \infty } \frac{\Sigma_S^2(T, k)}{|k|} = \frac{2}{T},
\end{equation}
which means that the wing behavior of implied volatility for the asset options is 
the sharpest possible one by \cite[Theorem 3.2 and 3.4]{L04}.

In the following of this subsection, we study the probability tails of $S$  which allows to replace the ``$\limsup$'' by the usual limit in \eqref{Lee formula} for the left wing of the asset options.
The next technical lemma, whose proof is postponed to Appendix,  shows that the extremal behavior of $V$ is mainly  due to one  large jump of the driving processes $Z$. 
\begin{Lem}\label{extremal behavior of V}
Fix $T>0$ and consider the variance process $V$ defined by (\ref{alpha Heston-root}).  Then there exists a nonzero boundedly finite measure $\delta$ on $\mathscr{B}(\bar{D}_0[0,T])$ with $\delta(\bar{D}_0[0,T]\backslash D[0,T])=0$ such that, as $u\rightarrow\infty$,
\begin{equation}\label{functional extremal behavior}u^{\alpha}\mathbb P({V}/u\in\cdot)\overset{\hat{w}}{\longrightarrow}\delta(\cdot)\quad on\quad \mathscr{B}(\bar{D}_0([0,T]),
\end{equation}
where  $\delta$ is given by:
\beqnn
\delta(\cdot)=\sigma_N^\alpha\int_0^T\big(b(1-e^{-as})+xe^{-as}\big)\int_0^\infty \mathbb E\Big[1_{\big\{w_t:=e^{-a(t-s)}y1_{[s,T]}(t)\in\;\cdot\;\big\}}\Big]\nu_{\alpha}(dy)ds,
\eeqnn
and $\nu_{\alpha}$ is defined by (\ref{Levymeasure}). We refer to Hult and Lindskog \cite[page 312]{HL07}
for the definition of $\bar{\mathbb{D}}_0[0,T]$ and the vague convergence $ \overset{\rm \hat{w}}{\longrightarrow}$. 

\end{Lem}

\begin{Pro}\label{prop1.3} Fix $t>0$. For any $x\ge 0$, we have that 
 \beqlb\label{tail prob of S}
 \mathbb P_x(-\log S_t>u)\sim-\Big(\frac{\sigma_N}{2a}\Big)^\alpha
\frac{\iota_\alpha(t)}{\alpha\cos(\pi\alpha/2)\Gamma(-\alpha)}u^{-\alpha},\quad u\rightarrow+\infty,
 \eeqlb
 where
 \beqnn
 \iota_\alpha(t)=e^{-\alpha at}\int_0^t(b(1-e^{-as})+xe^{-as})(e^{at}-e^{as})^\alpha ds.
 \eeqnn
 \end{Pro}
\proof
We have by (\ref{alpha Heston-root})  that 
\beqlb
\label{logS_t}
\log S_t=\log s_0+\int_0^t(r-\frac{1}{2}V_s)ds+\int_0^t\sqrt{V_s}dB_s.
\eeqlb
For any $t>0$, consider the asymptotic behavior of the probability tail for $\int_0^tV_sds$, that is,  $\mathbb P_x(\frac{1}{2}\int_0^tV_sds>x)$.
By Lemma \ref{extremal behavior of V}, as $u\rightarrow+\infty$,
\beqnn u^{\alpha}\mathbb P({V}/u\in\cdot)\overset{\hat{w}}{\longrightarrow}\delta(\cdot)\quad on\quad \mathscr{B}(\bar{D}_0[0,t]),
\eeqnn
Define the functional $h: \bar{D}_0[0,t]\longrightarrow \mathbb{R}_+$ by $h(w)=\frac{1}{2}\int_0^t
w_sds$. Let Disc($h$) be the set of discontinuities of $h$. By the definition of $h$ by (\ref{functional extremal behavior}), it is easy to see that $\delta({\rm Disc}(h)) =0$. It follows from \cite[Theorem 2.1]{HL07} that as $u\rightarrow+\infty$,
\beqnn
u^{\alpha}\mathbb P_x\Big(\frac{1}{2u}\int_0^tV_sds\in\cdot\Big)\overset{v}{\longrightarrow}\delta\circ h^{-1}(\cdot)\quad on\quad \mathscr{B}(\mathbb{R}_+),
\eeqnn
and 
\beqnn
\delta\circ h^{-1}(\cdot)
&=&
\sigma_N^\alpha\int_0^t
\mathbb E[V_s]\int_0^\infty1_{\{\frac{y}{2}\int_s^te^{-a(\zeta-s)}d\zeta\ \in\cdot\ \}}\nu_{\alpha}(dy)ds.
\eeqnn
Thus we have that
\beqnn
\mathbb P_x\Big(\frac{1}{2}\int_0^tV_sds>u\Big)\sim-\Big(\frac{\sigma_N}{2a}\Big)^\alpha
\frac{\iota_\alpha(t)}{\alpha\cos(\pi\alpha/2)\Gamma(-\alpha)}u^{-\alpha},\quad u\rightarrow+\infty.
\eeqnn
Furthermore we note that 
\beqnn
\mathbb E_x\Big[\Big(\int_0^t\sqrt{V_s}dB_s\Big)^2\Big]=\int_0^t\mathbb E_x[V_s]ds<\infty.
\eeqnn
In view of (\ref{logS_t}), we have that 
\beqnn
\mathbb P_x(-\log S_t>u)\sim \mathbb P_x\Big(\frac{1}{2}\int_0^tV_sds>u\Big),\quad u\rightarrow+\infty.
\eeqnn
\finproof
\begin{Cor}\label{prop1.4}  Let $\Sigma_S(T, k)$ be the implied volatility of the option written on the stock price $S$ with maturity $T$ and strike $K = e^k$. 
Then the left wing of $\Sigma_S(T, k)$ has the following
asymptotic shape as $k\rightarrow-\infty$:
\beqlb\label{eq-left-wing of S}
\frac{\sqrt{T}\Sigma_S(T, k)}{\sqrt{2}} &= &\sqrt{-k+\alpha\log(-k)-\frac{1}{2}\log\log(-k)}\nonumber\\
&&-\sqrt{\alpha\log(-k)-\frac{1}{2}\log\log(-k)}+O((\log(-k))^{-1/2}).
\eeqlb

\end{Cor}	
\proof Without loss of generality we assume $k<0$. Note that the put option price can be written as  
\beqnn
P(e^k):=\mathbb E[(e^k-S_T)_+]=\int_{-k}^\infty \mathbb P_x(-\log S_T>u)e^{-u}du.
\eeqnn
By Proposition \ref{prop1.3}, it is not hard to see that 
\beqnn
P(e^k)\sim-\Big(\frac{\sigma_N}{2a}\Big)^\alpha
\frac{\iota_\alpha(t)}{\alpha\cos(\pi\alpha/2)\Gamma(-\alpha)}e^k k^{-\alpha},\quad k\rightarrow-\infty.
\eeqnn
Then (\ref{eq-left-wing of S}) follows from the above asymptotic equality and \cite[Theorem 3.7]{Guli2010}. \finproof

\begin{figure}[h]
\caption{Implied volatilites for asset options }
\label{fig:impvol asset}
\begin{center}
\includegraphics[width=0.65\textwidth]{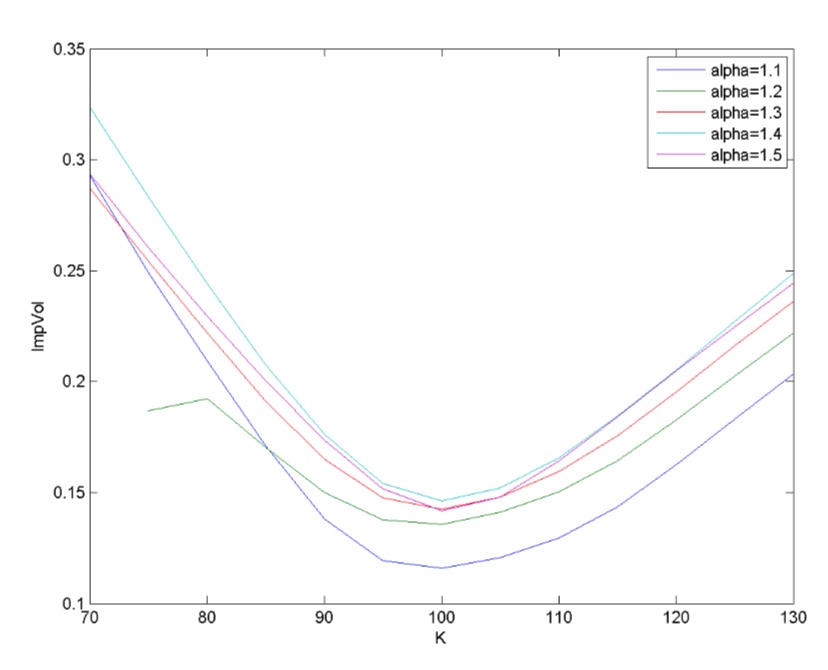}
\end{center}
\end{figure}

Figure \ref{fig:impvol asset} presents the implied volatility curves of the asset options.
We use a Monte Carlo method with $10^5$ trajectories with the  parameters $V_0= 0.0332$, $a=5$, $b=0.144$, $\sigma = 0.08$ and $\sigma_N = 1$ coherent with the ones of Nicolato et al. \cite{NPS2017}

%\begin{remark}[Self exciting effect]

%\end{remark}}

\subsection{Variance options}
We now consider the volatility and variance options  for which a large growing literature has been developed (see for instance \cite{GM2013}, \cite{NPS2017} and  \cite{Sepp2008}).
In particular, it is highlighted in \cite{Sepp2008} and  \cite{NPS2017}  about the upward-sloping implied volatility skew of VIX options. 
In the following, we derive  the asymptotic behavior of tail probability of $V$, which will imply the moment explosion condition for $V$ and the extreme behaviors of the variance options.
We begin by giving two technical lemmas. 

\begin{Lem} \label{Lemma 1.1} Let $X$ be a positive random variable.
\begin{enumerate}[(i)] 
\item (Karamata Tauberian Theorem \cite[Theorem 1.7.1]{BGT87}) For constants $C>0$, $\beta>0$ and a slowly varying function (at infinity) $L$,
\beqnn
\mathbb E[e^{-\lambda X}]\sim C\lambda^{-\beta}L(\lambda),\quad\mbox{as}\quad \lambda\rightarrow\infty,
\eeqnn
if and only if
\beqnn
\mathbb P(X\leq u)\sim\frac{C}{\Gamma(1+\beta)}u^\beta L(1/u),\quad\mbox{as}\quad u\rightarrow 0^+.
\eeqnn
\item (de Bruijn's Tauberian Theorem \cite[Theorem 4]{BT75}) Let $0\leq\beta\leq1$ be a constant, $L$ be a slowly varying function at infinity, and $L^*$ be the 
conjugate slowly varying function to $L$. Then
\beqnn
\log \mathbb E[e^{-\lambda X}]\sim-\lambda^\beta/L(\lambda)^{1-\beta} \quad \mbox{as} \ \lambda\rightarrow\infty,
\eeqnn
if and only if 
\beqnn
\log \mathbb P(X\leq u)\sim-(1-\beta)\beta^{\beta/(1-\beta)}u^{-\beta/(1-\beta)}L^*(u^{-1/(1-\beta)})\quad\mbox{as}\ u\rightarrow0^+.
\eeqnn
\end{enumerate}
\end{Lem}

\begin{Lem}\label{moments of V}  For any  $0<\beta<\alpha$,  there exists a locally bounded function $ C(\cdot)\ge 0$ such that for any $T\geq0$,
\beqnn
\mathbb {E}_x\Big[\sup_{0\le t\le T}V_t^\beta\Big]\le C(T)(1+x^\beta).
\eeqnn \end{Lem}

\begin{Pro}\label{prop1.2} (probability tails of $V_t$) Fix $t>0$. For any $x\ge 0$, we have that 
 \beqlb\label{tail prob of V}
\mathbb P_x(V_t>u)
 \sim
-\frac{\sigma_N^\alpha}{\alpha\Gamma(-\alpha)\cos(\pi\alpha/2)} \big(q_\alpha(t)+p_\alpha(t)x\big)
u^{-\alpha}, \quad \mbox{as}\quad u\to \infty, \eeqlb
where
 \beqnn\label{3.2}
&&~ p_\alpha(t) = \frac{1}{a(\alpha-1)}\left(e^{-a t} - e^{-\alpha a t}\right),
 \quad
q_\alpha(t) = b\left(\frac{1}{\alpha a}(1 - e^{-\alpha a t}) -
p_\alpha(t)\right).
 \eeqnn
Furthermore, \begin{enumerate}[(i)]
\item if $\sigma>0$, then
 \beqlb\label{small prob V with sigma}
\mathbb P_x(V_t\leq u)\sim u^{2ab/\sigma^2}\frac{\bar{v}_t^{2ab/\sigma^2}}{\Gamma\left(1+{2ab}/{\sigma^2}\right)}\exp\Big(-x\bar{v}_t-ab\int_{\bar{v}_t}^\infty \Big(\frac{z}{\Psi_\alpha(z)}-\frac{2}{\sigma^2z}\Big)dz\Big), \, \mbox{as}\,\, u\to 0,\eeqlb
where $\bar{v}_t$ is the minimal solution of the ODE 
\beqlb\label{ODE barV}
\frac{d}{dt}\bar{v}_t=-\Psi_\alpha(\bar{v}_t),\quad t>0,
\eeqlb
with singular initial condition $\bar{v}_{0+}=\infty$; %, and \beqnn\Psi_\alpha(q)=R(0,-q)=\frac{1}{2}\sigma^2q^2+aq-\frac{\sigma_N^\alpha}{\cos(\pi\alpha/2)}q^\alpha, \quad\mbox{for}\ q\geq0.\eeqnn
\item if $\sigma=0$, then 
\beqlb\label{small proba of V 2}
\log \mathbb P_x(V_t\leq u)\sim-\frac{\alpha-1}{2-\alpha}\left(-ab\cos\big(\frac{\pi\alpha}{2}\big)\right)^{\frac{1}{\alpha-1}}\sigma_N^{-\frac{\alpha}{\alpha-1}}u^{-\frac{2-\alpha}{\alpha-1}},\quad\mbox{as}\quad u\rightarrow 0.
\eeqlb
\end{enumerate}
\end{Pro}

\proof We have by  (\ref{alpha Heston-root})  that 
\beqlb\label{Ito transform}
V_t=e^{-at}V_0+ab\int_0^te^{-a(t-s)}ds+\sigma\int_0^te^{-a(t-s)}\sqrt{V_s}dB_s+\sigma_N\int_0^te^{-a(t-s)}V_{s-}^{1/\alpha}dZ_s.
\eeqlb Note that $\mathbb E_x[V_t]=e^{-at}x+b(1-e^{-at})$.  By Markov's inequality, 
\beqlb
\mathbb P_x\Big(\Big|\int_0^te^{-a(t-s)}\sqrt{V_s}dB_s\Big|>u\Big)&\leq&u^{-2}\mathbb E_x\Big[\int_0^te^{-2a(t-s)}V_sds\Big]\nonumber\\
&\leq& \Big(\frac{x}{a}+bt\Big)u^{-2}.\label{variance estimates}
\eeqlb
It follows from Lemma \ref{moments of V} that $\mathbb {E}[\sup_{0\leq t\leq T}(\sqrt[\alpha]{V_t})^{\alpha+\delta}]< \infty$ for 
$0<\delta<\alpha(\alpha-1)$. Then by  Hult and Lindskog \cite[Theorem~3.4]{HL07},
we have as $u\rightarrow\infty$, 
\beqlb
\mathbb P_x\Big(\sigma_N\int_0^te^{-a(t-s)}V_{s-}^{1/\alpha} dZ_s> u\Big)
 &\sim&
 \nu_{\alpha}(u,\infty)\sigma_N^{\alpha}\int_0^te^{-\alpha a(t-s)} \mathbb{E}_x[V_s]
ds\nonumber\\
&\sim&
-\frac{\sigma_N^\alpha}{\alpha\cos(\pi\alpha/2)\Gamma(-\alpha)}\big(q_\alpha(t) + p_\alpha(t)x\big)u^{-\alpha}.\label{stable tail estimates}
\eeqlb
In view of (\ref{Ito transform}), (\ref{variance estimates}) and (\ref{stable tail estimates}), the extremal behavior of $V_t$ is
determined by the forth term on the right-hand side of (\ref{Ito transform}).  Then we have, as $u\rightarrow \infty$,
 \beqnn
\mathbb{P}_x(V_t>u)
 &\sim&
\mathbb P_x\Big(\sigma_N\int_0^te^{-a(t-s)}V_{s-}^{1/\alpha} dZ_s> u\Big),
 \eeqnn
which gives (\ref{tail prob of V}).  On the other hand, by Proposition \ref{Pro: joint laplace transform}  we have 
\beqnn
\mathbb E_x\left[e^{-\lambda V_t}\right]=\exp\Big(-xv_t(\lambda)-ab\int_0^tv_s(\lambda)ds \Big),
\eeqnn
where $v_t(\lambda)$ is the unique solution of the following ODE:
\beqlb\label{ODE V}
\frac{\partial v_t(\lambda)}{\partial t}=-\Psi_\alpha(v_t(\lambda)),\quad v_0(\lambda)=\lambda.
\eeqlb
%Here $\Psi_\alpha(q)=R(0,-q)=\frac{1}{2}\sigma^2q^2+aq-\frac{\sigma_N^\alpha}{\cos(\pi\alpha/2)}q^\alpha$ for $q\geq0$. We refer to \cite[Theorem 3.1]{DawsonLi} for the LS of $V_t$; {\color{red}{SEE ALSO \cite[Proposition 3.2]{JMS2017}}}. 
It follows from \cite[Theorem 3.5, 3.8, Corollary 3.11]{Li11} that $\bar{v}_t=\uparrow\lim_{\lambda\rightarrow\infty}v_t(\lambda)$ exists in $(0,\infty)$
for all $t>0$, and $\bar{v}_t$ is the minimal solution of the singular initial value problem (\ref{ODE barV}). 

First consider the case of $\sigma>0$. By (\ref{ODE V}), 
\beqnn
\int_0^tv_s(\lambda)ds=\int_{v_t(\lambda)}^\lambda \frac{u}{\Psi_\alpha(u)}du=\int_{v_t(\lambda)}^\lambda \frac{2}{\sigma^2u}du+
\int_{v_t(\lambda)}^\lambda\Big(\frac{u}{\Psi_\alpha(u)}-\frac{2}{\sigma^2u}\Big)du, \quad \lambda>0,\ t>0.
\eeqnn
Note that $\frac{2}{\sigma^2u}-\frac{u}{\Psi_\alpha(u)}=O(u^{-(3-\alpha)})$ as $u\rightarrow\infty$ and 
thus $0<\int_{\bar{v}_t}^\infty\Big(\frac{2}{\sigma^2u}-\frac{u}{\Psi_\alpha(u)}\Big)du<\infty$. A simple calculation shows that 
\beqnn
\mathbb{E}_x \left[e^{-\lambda V_t}\right]
\sim\bar{v}_t^{2ab/\sigma^2}\lambda^{-2ab/\sigma^2}
\exp\left(-x\bar{v}_t-ab\int_{\bar{v}_t}^\infty \Big(\frac{u}{\Psi_\alpha(u)}-\frac{2}{\sigma^2u}\Big)du\right),
\quad \lambda\rightarrow0.
\eeqnn
Then Karamata Tauberian Theorem (see Lemma \ref{Lemma 1.1} (i)) gives (\ref{small prob V with sigma}). 

Now we turn to the case of $\sigma=0$. 
Denote  by $\sigma_1=-\frac{\sigma_N^\alpha}{\cos(\pi\alpha/2)}$. Recall that $\bar{v}_t=\uparrow\lim_{\lambda\rightarrow\infty}v_t(\lambda)\in(0,\infty),$ which is the minimal solution of the singular initial value problem (\ref{ODE barV}) with $\sigma=0$. 
Still by (\ref{ODE V}),
\beqnn
\log \mathbb{E}_x\left[e^{-\lambda V_t}\right]=
-xv_t(\lambda)-ab\int_{v_t(\lambda)}^\lambda \frac{1}{a+\sigma_1\lambda^{\alpha-1}}du\sim
\frac{ab}{\alpha-2}\frac{\lambda}{a+\sigma_1\lambda^{\alpha-1}}\sim\frac{ab}{\sigma_1(\alpha-2)}\lambda^{2-\alpha}.
\eeqnn
Then de Bruijn's Tauberian Theorem (see Lemma \ref{Lemma 1.1} (ii)) gives (\ref{small proba of V 2}).
\finproof

\begin{Cor}As a consequence of Proposition \ref{prop1.2}, we have, for any $\alpha\in(1,2)$, 
\begin{equation}
\{p\in\mathbb R: \mathbb E[V_t^p]<\infty\}=\big(-\frac{2ab}{\sigma^2}, \alpha\big)
%\sup\{p>0: E[V_t^p]<\infty\}=\alpha\quad\mbox{ and }\quad\sup\{p>0: E[V_t^{-p}]<\infty\}=\frac{2ab}{\sigma^2}
\end{equation}
where by convention $2ab/\sigma^2=+\infty$ if $\sigma=0$.
\end{Cor}
\proof By integration by parts, we have, for $p>0$, 
\[\mathbb E[V_t^{p}]=-\lim_{u\rightarrow\infty}u^p\mathbb P(V_t>u)+p\int_0^\infty u^{p-1}\mathbb P(V_t>u)du.\]By  Proposition \ref{prop1.2}, $\mathbb P(V_t>u)\sim C(t)u^{-\alpha}$ as $u\rightarrow\infty$ for some function $C(t)$. Then we obtain $\mathbb E[V_t^p]<\infty$ for $0\leq p<\alpha$ and $\mathbb E[V_t^p]=\infty$ for $p\geq\alpha$. Similarly, we consider $\mathbb E[(1/V_t)^p]$ and have $\mathbb P(1/V_t>u)\sim D(t)u^{-2ab/\sigma^2}$ as $u\rightarrow \infty$. Then we obtain $\mathbb E[(1/V_t)^p]<\infty$ for $0\leq p<2ab/\sigma^2$ and $\mathbb E[(1/V_t)^p]=\infty$ if $p\geq 2ab/\sigma^2$.
\finproof

\begin{Cor}\label{implied volatility based on variance processes} Let $\Sigma_V(T, k)$ be the implied volatility of call option written on the variance process $V$ with maturity $T$ and strike $K = e^k$ and let $\psi(q)=2 -4(\sqrt{q^2+q}-q)$. 
Then the right wing of $\Sigma_V(T, k)$ has the following
asymptotic shape:
\begin{equation}\label{eq-right-wing}
 \Sigma_V(T, k)\sim\Big(\frac{\psi(\alpha)}{T}\Big)^{1/2}\sqrt{k}, \quad k\rightarrow+\infty
\end{equation}
The left wing satisfies \begin{enumerate}[(i)]
\item if $\sigma>0$, then 
\begin{equation}\label{eq-left-wing 1}
 \Sigma_V(T, k)\sim\Big(\frac{\psi(\frac{2ab}{\sigma^2})}{T}\Big)^{1/2}\sqrt{-k}, \quad k\rightarrow-\infty;
\end{equation}
\item  if $\sigma=0$, then 
\begin{equation}\label{eq-left-wing 2}
 \Sigma_V(T, k)\sim\frac{1}{\sqrt{2T}}(-k)\Big(\log\frac{e^k}{P(e^k)}\Big)^{1/2}, \quad k\rightarrow-\infty.
\end{equation}
where $P(e^k)=E[(e^k-V_T)^+]$.
\end{enumerate}
\end{Cor}

\proof  Combining  (\ref{tail prob of V}) and \cite[Proposition 2.2-(a)]{NPS2017}, we obtain directly  (\ref{eq-right-wing}). Similarly, (\ref{small prob V with sigma}) and
\cite[Proposition 2.4-(a)]{NPS2017} leads to (\ref{eq-left-wing 1}). In the case where  $\sigma=0$, (\ref{small proba of V 2}) implies that $\sup\{p>0: \mathbb E[V_t^{-p}]<\infty\}=\infty$. Then (\ref{eq-left-wing 2}) follows from \cite[Theorem 2.3-(iii)]{NPS2017}.
\finproof

Corollary \ref{implied volatility based on variance processes} gives the explicit behavior of the implied volatility of variance options with extreme strikes far from the moneyness. We note that 
the right wing depends only on the parameter $\alpha$ which is the characteristic parameter of the jump term. When $\alpha$
decreases, the tail becomes heavier and the slope in  (\ref{eq-right-wing}) increases. In contrast, the left wing depends on the parameters which belong to the pure  CIR part with Brownian diffusion and the explaining coefficient $2ab/\sigma^2$ in \eqref{eq-left-wing 1} is linked to the Feller condition. When the Brownian term disappears, i.e. $\sigma=0$, then there occurs a  discontinuity on the left wing behavior of the variance volatility surface.

\section{Jump cluster behaviour }\label{sec: clusters}

In this section, we study the jump cluster phenomenon by giving a  decomposition formula of the variance process $V$ and we analyze some properties of the cluster processes. 

\subsection{Cluster decomposition of the variance process}

Let us fix a jump threshold ${y}=\sigma_Z\overline{y}$ and denote by $\{\tau_n\}_{n\geq 1}$ the sequence of jump times of $V$ whose sizes are larger than $y$.  We call $\{\tau_n\}_{n\geq 1}$ the large jumps.
By separating the large  and small jumps, the variance process \eqref{Vol integral} can be written as 
\begin{equation}\label{V integral rewritten}
\begin{array}{rcl}
\displaystyle V_t & \displaystyle = V_0 &\displaystyle + \int_0^t a\left(b-\frac{\sigma_N\Theta(\alpha, y)V_s}{a}-V_s\right)  ds + 
 \sigma\int_{0}^t \int_0^{V_s} W(ds,du) \\
& & \displaystyle +\sigma_N \int_{0}^t \int_0^{V_{s-} } \int_0^{\overline{y}}\zeta \widetilde{N}
(ds, du, d\zeta)++\sigma_N \int_{0}^t \int_0^{V_{s-} } \int_{\overline{y}}^{\infty}\zeta {N}
(ds, du, d\zeta)
\end{array}
\end{equation}
where
\begin{eqnarray}\label{def-a-b-theta}
\Theta(\alpha, y) &=& \int_{\overline{y}}^{\infty}\zeta\nu_{\alpha}(d\zeta)=\frac{2}{\pi}\alpha\Gamma(\alpha-1)\sin\left(\frac{\pi\alpha}{2}\right)\overline{y}^{1-\alpha}.
\end{eqnarray}
We denote by \[\widetilde{a}(\alpha, y) = a+ \sigma_N\Theta(\alpha, y) \quad \text{ and }\quad 
\widetilde{b}(\alpha, y) = \frac{ab}{a+ \sigma_N \Theta(\alpha, y)}.\]
Then between two large jumps times, that is, for any $t\in[\tau_n, \tau_{n+1})$, we have 
\begin{equation}\label{lambda integral-without-compensation}
\begin{array}{rcl}
\displaystyle V_t & \displaystyle = V_{\tau_n} &\displaystyle + \int_{\tau_n}^t \widetilde{a}(\alpha, y)  \big(\widetilde{b}(\alpha, y)   - V_s\big)ds + 
 \sigma\int_{\tau_n}^t \int_0^{V_s} W(ds,du) \\
& & \displaystyle +\sigma_N \int_{\tau_n}^t \int_0^{V_{s-} } \int_0^{\overline{y}}\zeta \widetilde{N}
(ds, du, d\zeta).
\end{array}
\end{equation}
The expression \eqref{lambda integral-without-compensation} shows that two phenomena arise between two large jumps. First, the mean long-term level $b$ is reduced.
This effect is standard %and appears also \textcolor{red}{in a L\'evy-Ornstein-Uhlenbeck model}, 
since the mean level $\widetilde{b}(\alpha, y)$ becomes lower  to compensate  the large jumps in order to preserve the global mean level $b$.
Second and more surprisingly, the mean reverting speed $a$ is augmented. That is, the volatility decays more quickly between two jumps.
Moreover, this speed is greater when the parameter $\alpha$ decreases and tends to infinity as $\alpha$  approaches $1$ since  $\Theta(\alpha, y) \sim (\alpha-1)^{-1}$.

We introduce the truncated process of $V$ up to the jump threshold, which will serve as the fundamental part in the decomposition, as
\begin{equation}\label{rhat1}\begin{split}
{V}^{(y)}_t =V_0 &+ \int_0^t \widetilde{a}(\alpha, y) \big( \widetilde{b}(\alpha, y)  - {V}_s^{(y)}  \big) ds 
+ \sigma \int_0^t \int_0^{{V}_s^{(y)}} W(ds,du)\\
&+ \sigma_N \int_0^t \int_0^{{V}_{s-}^{(y)} } \int_0^{\overline{y}}\zeta \widetilde{N}
(ds,du,d\zeta), \quad t\geq 0.
\end{split}\end{equation}
Similar as  $V$, the process $V^{(y)}$ is also a CBI process. By definition, the jumps of the process $V^{(y)}$ are all smaller than $y$. In addition, $V^{(y)}$ coincides with $V$ before the first large jump $\tau_1$. The next result studies the first large jump and its jump size, which will be useful for the decomposition.
\begin{Lem}\label{large jump size}  We have 
\beqlb\label{tau1}
\mathbb{P}(\tau_1>t)=\mathbb{E}\Big[\exp{\Big\{-\Big(\int_{\overline y}^\infty\mu_\alpha(d\zeta)\Big)\Big(\int_0^t{V}^{(y)}_sds\Big)\Big\}}\Big].
\eeqlb
The jump  $\Delta V_{\tau_1}:=V_{\tau_1}-V_{\tau_1-}$ is independent of
$\tau_1$ and $V^{(y)}$, and satisfies 
\beqlb\label{size distribution}
\mathbb{P}(\Delta V_{\tau_1}\in d\zeta)=1_{\{\zeta>{y}\}}\,\frac{ \alpha{y}^\alpha} {\zeta^{1+\alpha}}d\zeta.
\eeqlb
\end{Lem}

It is not hard to 
see that $\mathbb{P}(V_t\geq V^{(y)}_t,\forall t\geq0)=1$. Then the large jump in \eqref{V integral rewritten} can be separated into two parts as
\beqlb\label{large jumps}
\int_0^t\int_0^{V_{s-}}\int_{\bar{y}}^\infty N(ds,du,d\zeta)=\int_0^t\int_0^{V^{(y)}_{s-}}\int_{\bar{y}}^\infty N(ds,du,d\zeta)+\int_0^t\int_{V^{(y)}_{s-}}^{V_{s-}}\int_{\bar{y}}^\infty N(ds,du,d\zeta).
\eeqlb
Let
\beqlb\label{mother jump intensity}
J_t^{(y)}=\int_0^t\int_0^{V^{(y)}_{s-}}\int_{\bar{y}}^\infty N(ds,du,d\zeta),\quad t\geq 0\eeqlb
which is a point process whose arrival times $\{T_n\}_{n\geq 1}$ coincide with part of the large jump times and those jumps are called the mother jumps. By definition, the mother jumps form a subset of large jumps. Each mother jump will induce a cluster process $v^{(n)}$ which starts from time $T_n$  with initial value $\Delta V_{T_n}=V_{T_n}-V_{T_n-}$ and is given recursively by  
\begin{equation}\label{cluster}\begin{split}
v_t^{(n)}=\Delta V_{T_n}  &-a \int_{T_n}^t v^{(n)}_s ds + \sigma \int_{T_n}^t 
\int_{V^{(y)}_s+\sum_{i=1}^{n-1}v^{(i)}_s}^{V^{(y)}_s+\sum_{i=1}^{n}v^{(i)}_s} W(ds, du)\\
 &+ \sigma_Z \int_{T_n}^t \int_{V^{(y)}_{s-}+\sum_{i=1}^{n-1}v^{(i)}_{s-}}^{V^{(y)}_{s-}+\sum_{i=1}^{n}v^{(i)}_{s-}} \int_{\mathbb{R}^+} 
\zeta \widetilde{N}(ds,du, d\zeta),\quad t\in[T_n,\infty).\end{split}\end{equation}

The next result provides the decomposition of $V$ as the sum of  the fundamental process $V^{(y)}$ and  a sequence of cluster processes. The decomposition form is inspired by Duquesne and Labbe \cite{DuqLab}. %where they establish an excursion representation for a CB process with finite variation.

%We distinguish so called mother jumps from large jumps of $V$, which are defined by the sequence of arrival times $\{T_n\}_{n\geq 1}$ of the point process $\{J_t^{(y)}: t\geq0\}$. A careful investigation (see the proof of Proposition \ref{Pro: decomposition}) shows that, between two mother jump time, $V$ satisfies \beqlb\label{n interval}V_t=V^{(y)}_t+\sum_{k=1}^nv^{(k)}_t,\quad t\in[T_n,T_{n+1}).\eeqlb

%Intuitively, for each $n$, the mother jump $\Delta_{T_n}$ induces a new cluster process $v^{(n)}$ starting from time $T_n$ with initial value $\Delta_{T_n}$, which gives the jump clustering part. 
\begin{Pro}\label{Pro: decomposition} 
The variance process $V$ given by \eqref{Vol integral} has the decomposition:
\begin{equation}\label{decomposition}{V_t=V^{(y)}_t+\sum_{n=1}^{J^{(y)}_t}u_{t-T_n}^{(n)}},\quad t\geq 0, \end{equation}
where $u^{(n)}_t=v^{(n)}_{T_n+t}$ with $v^{(n)}$ given by (\ref{cluster}).
Moreover, we have that 
\begin{enumerate}[(1)]
\item $\{u^{(n)}: n=1,2,\cdots\}$ is the sequence of independent identically distributed processes and for each $n$, $u^{(n)}$ has the same distribution as an $\alpha$-$\mathrm{CIR}(a,0,\sigma,\sigma_Z,\alpha)$ process given by
\beqlb\label{CB}
u_t=u_0-a\int_0^tu_sds  + \sigma\int_0^t  \sqrt{u_s} dB_s
+\sigma_N\int_0^t\sqrt[\alpha]{u_{s-}}dZ_s,
\eeqlb
where $u_0\overset{d}{=}\Delta V_{\tau_1}$ and its distribution is given by \eqref{size distribution}.  
\item  The pair $({V}^{(y)}, J^{(y)})$ is independent of $\{u^{(n)}\}$. Conditional on ${V}^{(y)}$, $J^{(y)}$ is a time inhomogenous Poisson process with intensity function $\big(\int_{\bar y}^\infty\nu_\alpha(d\zeta)\big){V}^{(y)}_\cdot$.
\end{enumerate}
\end{Pro}

Note that each cluster process has the same distribution as an $\alpha$-square root jump process which is similar to \eqref{alpha Heston-root} but with parameter $b=0$, that is, an $\alpha$-$\mathrm{CIR}(a,0,\sigma,\sigma_Z,\alpha)$ process also known as a CB process without immigration.  The jumps given by $(J_t,t\geq 0)$ are called mother jumps in the sense that each mother jump $T_n$ will induce a cluster of jumps, or so-called son jumps, via its cluster (branching) process $u^{(n)}$. Conversely, any jump from $\big(\int_0^t\int_{V^{(y)}_{s-}}^{V_{s-}}\int_{\bar{y}}^\infty N(ds,du,d\zeta), t\geq0\big)$ in \eqref{large jumps}, that is, a large jump but not mother jump, is a child jump of some mother jump.

%TO RELATED TO HAWKES PROCESS CONSTRUCTION (MORE DETAILS??)

%We denote by $D(\mathbb R_+)$ the space of c\`adl\`ag functions $\omega:\mathbb R_+\rightarrow\mathbb R_+$  with the Skorokhod topology.For any $x>0$, an  $\alpha$-$\mathrm{CIR}(a,0,\sigma,\sigma_Z,\alpha)$ process with initial value $x$ can be seen as a random variable on $(\Omega,\mathcal A,\mathbb Q)$ which takes values in $D$ and we denote by $\mathbb Q_x$ the probability law of this random variable, which is a Borel probability measure on $D$. We introduce a Poisson random measure  $M(dt,du,d\omega)$  on $\mathbb{R}_+^2\times D$ whose intensity measure is $dt\,du\int_{\overline{y}}^\infty\mathbb{Q}_x(d\omega)\nu_{\alpha}(dx),$ and which is independent of $W$ and $N$. By using this auxiliary random measure  on an extended probability space, we have the following equivalent form of (\ref{decomposition}): \beqnnV_t={V}^{(y)}_t+\int_0^t\int_0^{{V}^{(y)}_{s-}}\int_{D}\omega_{t-s}M(ds,du,d\omega),\eeqnn and thus\beqnnJ_t^{(y)}=\int_0^t\int_0^{V_{s-}^{(y)}}\int_{D}M(ds,du, d\omega).\eeqnn
 
\subsection{The cluster processes} 
 We finally focus on the cluster processes and present some of their properties. We are particularly interested in two quantities. The first one is the number of clusters before a given time $t$, which is equal to the number of mother jumps. The second one is the duration of each cluster process.
\begin{Pro}\label{Pro: decomposition2} \begin{enumerate}[(1)]
\item The expected number of clusters during  $[0,t]$ is
\begin{equation}\label{number cluster}\mathbb E[J_t^{(y)}]=
\frac{(1-\alpha)\sigma_Z^\alpha}{\cos(\pi\alpha/2)\Gamma(2-\alpha)y^\alpha}\Big(\widetilde{b}(\alpha, y)t+\frac{V_0-\widetilde{b}(\alpha, y)}{\widetilde{a}(\alpha, y)}(1-e^{-\widetilde{a}(\alpha, y)t})\Big).\end{equation}
\item Let $\theta_n:=\inf\{t\geq0: u_t^{(n)}=0\}$ be the duration of the cluster $u^{(n)}$.  We have 
$ \mathbb P(\theta_n<\infty)=1$  and 
\begin{equation}\label{duration cluster}\mathbb E[\theta_n]= \alpha y^\alpha\int_0^\infty\frac{dz}{\Psi_{\alpha}(z)}\int_{y}^{\infty}\frac{1-e^{-\zeta z}}{\zeta^{1+\alpha}}d\zeta.
\end{equation}
\end{enumerate}
\end{Pro}

We note that the expected duration of all clustering processes are equal, which means that the initial value of $u^{(i)}$, that is, the jump size of the triggering mother jump has no impact on the duration. By \eqref{duration cluster}, we have
\begin{equation*}\mathbb E[\theta_n]=
 \alpha \int_0^\infty\frac{dz}{\Psi_{\alpha}(z)}\int_{1}^{\infty}\frac{1-e^{-\zeta y z}}{\zeta^{1+\alpha}}d\zeta,
 \end{equation*}
 which implies that $\mathbb{E}[\theta_n]$ is increasing with $y$. It is natural as larger jumps induce longer-time effects. But typically, the duration time is short, which means that there
 is no long-range property for $\theta_n$, because we have the following estimates:
 \beqlb\label{light tails}
 \mathbb{P}(\theta_n>t)\leq\frac{\alpha y}{\alpha-1}q_1e^{-a(t-1)},\quad t>1,
 \eeqlb
for some constant $0<q_1<\infty$.

%We need wait a long time for a extremely large jump. However once such case happens,  more large jumps might be induced in a cluster duration. Although the duration is increasing with respect to $y$, it is always relatively short (no long-range) due to finite expectation and exponentially decreasing probability tails  given by (\ref{light tails}).

\begin{figure}
\caption{The expected number of clusters (left) and the duration of one cluster (right) as a function of the jump threshold $y$, for different values of $\alpha$.}
\begin{center}
\includegraphics[width=0.49\textwidth]{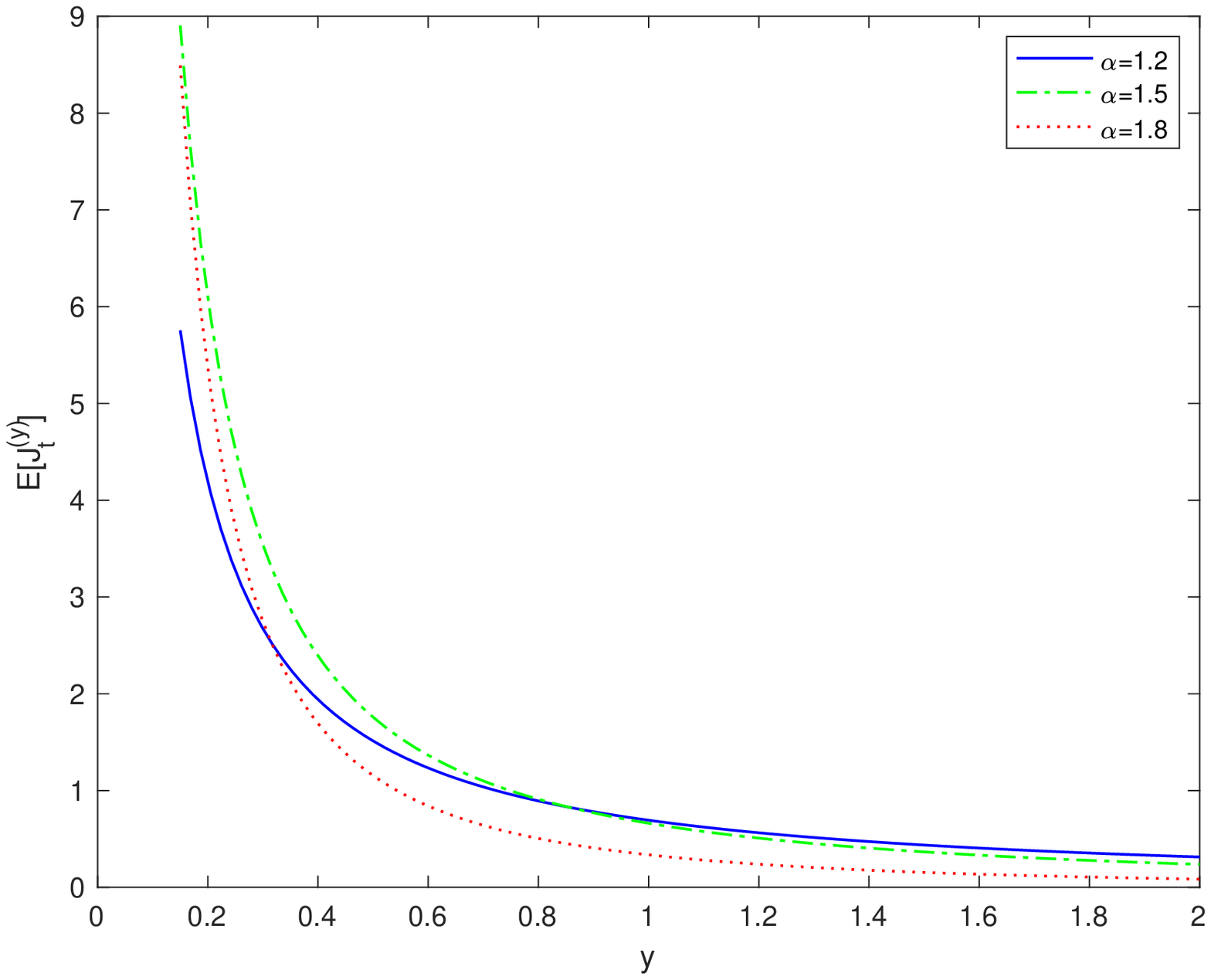}
\includegraphics[width=0.49\textwidth]{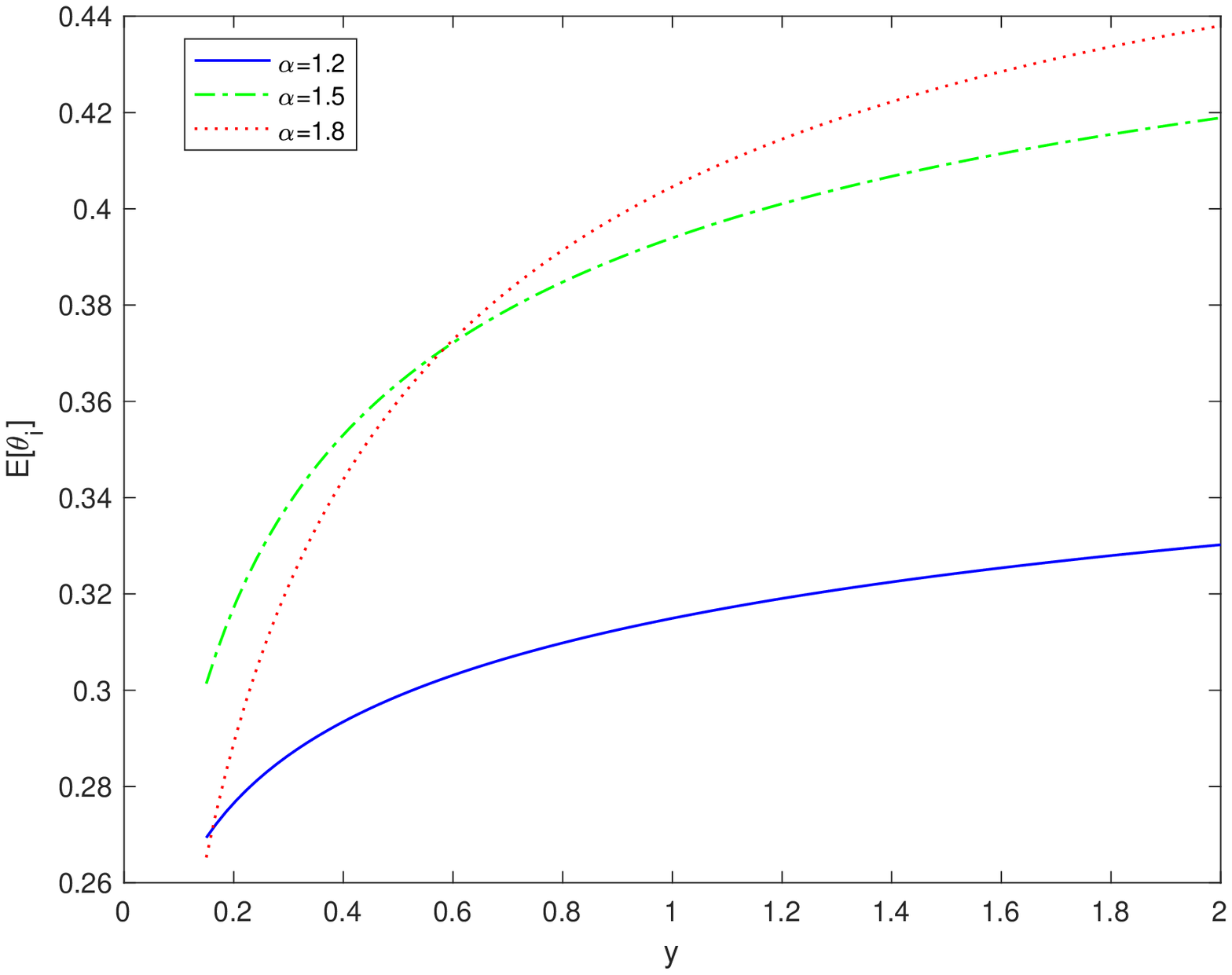}
\end{center}
\label{fig: cluster}       
\end{figure}

We illustrate in Figure \ref{fig: cluster} the behaviors of the jump cluster processes by the above proposition. The parameters are similar as in Figure \ref{fig: variance} except that we compare three different values for $\alpha=1.2$, $1.5$ and $1.8$. The first graph shows the expected number of clusters given by \eqref{number cluster}, as a function of $y$ for a period of $t=14$.  We see that when the jump threshold $y$ increases, there will be less clusters. In other words, we need to wait a longer time to have a very large mother jump. However once such case happens, more large son jumps might be induced during a cluster duration so that the duration is increasing with $y$. For large enough $y$, the number of clusters is decreasing with $\alpha$. In this case, the large jumps play a dominant role. For small values of $y$, there is a mixed impact of both small and large jumps which breaks down the monotonicity with $\alpha$.  The second graph illustrates the duration of one cluster which is given by \eqref{duration cluster}. %The duration time is increasing with $y$, and for large enough $y$, is also increasing with $\alpha$. 
Although the duration is increasing with respect to $y$, it is always relatively short due to finite expectation and exponentially decreasing probability tails 
given by (\ref{light tails}).

When the jump threshold $y$ becomes extremely large, the point process $\{J_t^{(y)}\}$ is asymptotic to a Poisson process and the expected number of clusters converges to a fixed level, as shown by the following result. %In other word,  each extremal large jump might happen at rate $\lambda n$. It seems reasonable since each (big) economy crisis happen periodically...  In the simulation we "possibly observe" that the number of clusters is always increasing and converges to a fixed level. To verify it, we consider the asymptotic behavior of $\{J_t^{(y)}\}$  for long time $t$ and large $y$.

\begin{Pro}\label{Pro: decomposition3}\; Let $\{y_n\}_{n\geq 1}$ be the sequence of positive thresholds with $y_n\sim cn^{1/\alpha}$ as $n\rightarrow\infty$ where  $c$  is some positive constant. Then for each $t\geq0$, 
\beqlb\label{convergence of J_t}
J^{(y_n)}_{nt}\overset{w}{\longrightarrow} J_t,
\eeqlb	
as $n\rightarrow\infty$, where $J$ is a Poisson process with the parameter $\lambda$ given by
\beqnn
\lambda=-\frac{\sigma_N^\alpha b}{\alpha\cos(\pi\alpha/2)\Gamma(-\alpha)c^\alpha}, 
\quad 1<\alpha<2.
\eeqnn
\end{Pro}

\section{Appendix}
%\subsection{Proofs for Section \ref{sec: affine}}

{\bf Proof of Proposition \ref{Pro: joint laplace transform}. } As a a direct consequence of \cite{DPS2000} and  \cite{KR11}, the proof mainly serves to provide the explicit form of the generalized Riccati equations.
By (\ref{Heston-integral})  we have 
\beqnn
dY_t=(r-\frac{1}{2}V_t)dt+\rho\int_0^{V_t}W(dt,du)+\sqrt{1-\rho^2}\int_0^{V_t}\overline{W}(dt,du).
\eeqnn
By Ito's formula,  we have that the process $(Y_t, V_t, \int_0^tV_sds)$ is an affine process with generator given by 
\beqnn
Af(y,v,u)&=& (r-\frac{1}{2}v)f'_y(y,v,u)+ a(b-v)f'_v(y,v,u)+vf'_u(y,v,u)\\
& &+\frac{1}{2}vf''_{yy}(y,v,u)+\rho\sigma vf''_{yv}(y,v,u)+\frac{1}{2}\sigma^2vf''_{vv}(y,v,u)\\
& &+\sigma_N^\alpha v\int_0^\infty \big(f(y,v+\zeta,u)-f(y,v,u)-f'_v(y,v,u)\zeta\big)\nu_{\alpha}(d\zeta).
\eeqnn

Denote by $X_t=(Y_t,V_t,\int_0^tV_sds)$. We aim to find some functions $(\phi,\tilde{\Psi})\in\mathbb{C}\times\mathbb{C}^3$
with $\phi(0,\xi)=0$ and $\tilde{\Psi}(0,\xi)=\xi$ such that the following duality holds
\beqlb\label{duality}
{\mathbb E}\left[e^{\langle\xi,X_T\rangle}\right]=\exp\Big(\phi(T,\xi)+\langle\tilde{\Psi}(T,\xi),X_0\rangle\Big).
\eeqlb
In fact, if 
\beqnn
M_t=f(t,X_t)=\exp\Big(\phi(T-t,\xi)+\langle\tilde{\Psi}(T-t,\xi),X_t\rangle\Big)
\eeqnn
is a martingale, then we immediately have that 
\beqnn
\mathbb{E}[e^{\langle \xi,X_T\rangle}]=\mathbb{E}[M_T]=M_0=\exp\Big(\phi(T,\xi)+\langle\tilde{\Psi}(T,\xi),X_0\rangle\Big),
\eeqnn
which implies (\ref{duality}). Now assume that  $(\phi,\tilde{\Psi})$ are sufficiently differential and applying the Ito formula to 
$f(t,X_t)$, we have that 
\beqnn
M_T-M_0&=&\mbox{local martingale part}-\int_0^Tf(t,X_t)\Big(\dot{\phi}(T-t,\xi)+\langle X_t,\dot{\tilde{\Psi}}(T-t,\xi)\rangle\Big)dt\\
&&+\int_0^Tf(t,X_t)\tilde{\Psi}_1(T-t,\xi)(r-\frac{1}{2}V_t)dt+\int_0^Tf(t,X_t)\tilde{\Psi}_2(T-t,\xi)a(b-V_t)dt\\
&&+\int_0^Tf(t,X_t)\tilde{\Psi}_3(T-t,\xi)V_tdt+\frac{1}{2}\int_0^T
f(t,X_t)\tilde{\Psi}^2_1(T-t,\xi)V_tdt\\
&&+\rho\sigma\int_0^Tf(t,X_t)\tilde{\Psi}_1(T-t,\xi)\tilde{\Psi}_2(T-t,\xi)V_tdt+\frac{1}{2}\sigma^2\int_0^Tf(t,X_t)\tilde{\Psi}_2^2(T-t,\xi)
V_tdt\\
&&+\sigma_N^{\alpha}\int_0^Tf(t,X_t)V_t\int_0^\infty \Big[\exp\{\tilde{\Psi}_2(T-t,\xi)z\}-1-\tilde{\Psi}_2(T-t,\xi)z\Big]\nu_{\alpha}(dz)
\eeqnn
where $\tilde{\Psi}=(\tilde{\Psi}_1,\tilde{\Psi}_2,\tilde{\Psi}_3)$. Then $f(t,X_t)$ is a local martingale, if 
\beqnn
\dot{\phi}(T-t,\xi)=r\tilde{\Psi}_1(T-t,\xi)+ab\tilde{\Psi}_2(T-t,\xi),\quad\dot{\tilde{\Psi}}_1(T-t,\xi)=0, \quad\dot{\tilde{\Psi}}_3(T-t,\xi)=0,
\eeqnn
and 
\beqnn
\dot{\tilde{\Psi}}_2(T-t,\xi)&=&-\frac{1}{2}\tilde{\Psi}_1(T-t,\xi)-a\tilde{\Psi}_2(T-t,\xi)+\tilde{\Psi}_3(T-t,\xi)\\
&&+\frac{1}{2}\tilde{\Psi}_1^2(T-t,\xi)+\rho\sigma\tilde{\Psi}_1(T-t,\xi)\tilde{\Psi}_2(T-t,\xi)+\frac{1}{2}\sigma^2\tilde{\Psi}^2_2(T-t,\xi)\\
&&+\sigma_N^{\alpha}\int_0^\infty \Big(e^{z\tilde{\Psi}_2(T-t,\xi)}-1-z\tilde{\Psi}_2(T-t,\xi))\Big)\nu_{\alpha}(dz).
\eeqnn
Then we have that $\tilde{\Psi}_1(t,\xi)=\xi_1$ and $\tilde{\Psi}_3(t,\xi)=\xi_3$ for $0\leq t\leq T$. Furthermore $\tilde{\Psi}_2(t,\xi)$ solves the ODE
\beqnn
\dot{\tilde{\Psi}}_2(t,\xi)&=&-\frac{1}{2}\xi_1-a\tilde{\Psi}_2(t,\xi)+\xi_3+\frac{1}{2}\xi_1^2+\rho\sigma\xi_1\tilde{\Psi}_2(t,\xi)+\frac{1}{2}\sigma^2\tilde{\Psi}^2_2(t,\xi)\\
&&+\sigma_N^{\alpha}\int_0^\infty \Big(e^{z\tilde{\Psi}_2(t,\xi)}-1-z\tilde{\Psi}_2(t,\xi))\Big)\nu_{\alpha}(dz)\\
&=&-\frac{1}{2}\xi_1-a\tilde{\Psi}_2(t,\xi)+\xi_3+\frac{1}{2}\xi_1^2+\rho\sigma\xi_1\tilde{\Psi}_2(t,\xi)+\frac{1}{2}\sigma^2\tilde{\Psi}^2_2(t,\xi)
-\frac{\sigma_N^\alpha}{\cos(\pi\alpha/2)}(-\tilde{\Psi}_2(t,\xi))^{\alpha}
\eeqnn
Now let $\Psi(t,\xi)=\tilde{\Psi}_2(t,\xi)$, which obviously satisfies the ODE (\ref{generalized Racci}) and $$\phi(t,\xi)=\int_0^t(r\xi_1+ab\Psi(s,\xi)ds.$$The proof is thus complete. 
\finproof

%\subsection{Proofs for Section \ref{sec: extreme strike}}

\noindent {\bf Proof of Lemma \ref{extremal behavior of V}. } Consider (\ref{Ito transform}). By Doob's inequality, 
$$
\mathbb E_x\Big[\sup_{0\leq t\leq T}\Big|\int_0^t e^{-a(t-s)}\sqrt{V_s}dB_s\Big|^2\Big]
\leq 4 \mathbb E_x \Big[\int_0^T e^{2as} V_s ds\Big] \leq \frac{2x+b}{2a}e^{2aT}，
$$
which implies that $u^\alpha \mathbb P_x(\sup_{0\leq t\leq T}|\int_0^t e^{-a(t-s)}\sqrt{V_s}dB_s|>u)\rightarrow0$ as $u\rightarrow\infty$.
Then, in view of (\ref{Ito transform}), the extremal behavior of $V_t$ in the sense of (\ref{functional extremal behavior}) is determined by
\beqnn
\sigma_N\int_0^te^{-a(t-s)}\sqrt[\alpha]{V_{s-}} dZ_s=e^{-at}\cdot \sigma_N\int_0^te^{as}\sqrt[\alpha]{V_{s-}}dZ_s:=X_t\cdot Y_t.
\eeqnn
Note that  $\mathbb{E}[\sup_{0\leq t\leq T}(\sqrt[\alpha]{V_t})^{\alpha+\delta}]< \infty$ for 
$0<\delta<\alpha(\alpha-1)$ from Lemma \ref{moments of V}. Then by \cite[Theorem 3.4]{HL07},
we have as $u\rightarrow\infty$, 
\begin{equation}\label{functional extremal behavior2}u^{\alpha}\mathbb P({Y}/u\in\cdot)\overset{\hat{w}}{\longrightarrow}\delta_Y(\cdot)\quad on\quad \mathscr{B}(\bar{D}_0([0,T]),
\end{equation}
where  $\delta$ is given by:
\beqnn
\delta_Y(\cdot)=T \mathbb E\Big[\int_0^\infty 1_{\{w_t:=\sigma_N e^{a\tau}\sqrt[\alpha]{V_\tau}y1_{[\tau,T]}(t)\in\cdot\}}\nu_{\alpha}(dy)\Big],
\eeqnn
where $\tau$ is uniformly distributed on $[0, T]$ and independent of $V$.
Furthermore, by \cite[Theorem 3.1]{HL07}, we have as $u\rightarrow\infty$, 
\beqnn\label{functional extremal behavior2}u^{\alpha}\mathbb P({XY}/u\in\cdot)\overset{\hat{w}}{\longrightarrow}\delta_Y(w\in\bar{D}_0[0,T]: Xw\in\cdot)\quad on\quad \mathscr{B}(\bar{D}_0([0,T]),
\eeqnn
A simple calculation shows that 
\beqnn
\delta(\cdot):=\delta_Y(w\in\bar{D}_0[0,T]: Xw\in\cdot)=\sigma_N^\alpha\int_0^T
\mathbb E[V_s]\int_0^\infty1_{\{w_t=e^{-a(t-s)}y1_{[s,T]}(t)\in\cdot\}}\nu_{\alpha}(dy)ds
\eeqnn
\finproof

\noindent{\bf Proof of Lemma \ref{moments of V}. } By \eqref{Ito transform}, an elementary inequality shows that there exists a locally bounded function $C_1(\cdot)$ such that
 \beqlb\label{supV}
\mathbb {E}_x\Big[\sup_{0\le t\le T}V_t^\beta\Big]
 &\leq&
C_1(T)\Big(x^\beta + b^{\beta} +\sigma^\beta \mathbb E_x\Big[\sup_{0\leq t\leq T}\big|\int_0^te^{-a(t-s)}\sqrt{V_s}dB_s\big|^\beta\Big]\nonumber\\
&&+\sigma^\beta_N \mathbb{E}_x\Big[\sup_{0\le t\le T} \big|\int_0^te^{-a(t-s)}V_{s-}^{1/\alpha} dZ_s\big|^\beta\Big]\Big).
 \eeqlb
By H\"older's inequality and Doob's martingale inequality, there exist a locally bounded function $C_2(\cdot)$ such that
\begin{equation}
\begin{split}
&\quad\mathbb E_x\Big[\sup_{0\leq t\leq T}\Big|\int_0^te^{-a(t-s)}\sqrt{V_s}dB_s\Big|^\beta\Big]\leq
2^\beta\mathbb{E}_x\Big[\Big(\int_0^Te^{2 as}V_s ds\Big)^{\beta/2} \Big]\nonumber\\
&\le
2^\beta\Big(\int_0^T\mathbb{E}_x[e^{2 as}V_s] ds\Big)^{\beta/2} \nonumber \le
C_2(T)\left(x^{\beta/2}e^{\beta aT/2} + e^{\beta aT}\right). \label{integral inequality 1}
\end{split} \end{equation}
Moreover, by  Long \cite[Lemma 2.4]{Lon10}, which is a generalization of Rosinski and Woyczynski \cite[Theorem 3.2]{RW85}, there exist locally bounded functions $C_3(\cdot)$ and $C_4(\cdot)$ such that
 
  \begin{equation}\begin{split}
&\quad\mathbb{E}_x\Big[\sup_{0\le t\le T}\Big|\int_0^t e^{-a(t-s)}
V_{s-}^{1/\alpha} dZ_s\Big|^\beta\Big]
 \leq
C_3(T)\mathbb{E}_x\Big[\Big(\int_0^T e^{\alpha as} V_s
ds\Big)^{\beta/\alpha}\Big]\nonumber\\
&\le
C_3(T)\Big(\int_0^T\mathbb{E}_x[e^{\alpha as}V_s] ds\Big)^{\beta/\alpha} \nonumber\le
C_4(T)\left(x^{\beta/\alpha}e^{\beta a(1-1/\alpha)T} + e^{\beta aT}\right).
\label{integral inequality 2}
 \end{split}\end{equation}
By combining (\ref{supV}), (\ref{integral inequality 1}) and (\ref{integral inequality 2}), we have the lemma.
\finproof

%\subsection{Proofs for Section \ref{sec: clusters}}

\noindent{\bf Proof of Lemma \ref{large jump size}.} %Recall that $y=\overline{y}/\sigma_{Z}$. (\ref{probability rep}) has been established in Jiao {\it et al.} (2017). 
%The proof  of (\ref{tau1}) is based on He and Li \cite[Theorem 3.2]{HL16}.  
By (\ref{Vol integral}), we note that 
\beqnn
\{\tau_1>t\}=\Big\{\tau_1>t, \ \int_0^t\int_0^{V_{s-}}\int_{\overline{y}}^\infty\zeta N(ds,du,d\zeta)=0\Big\}.
\eeqnn
Since $V^{(y)}$ coincides with $V$ up to 
$\tau_1$, the comparison between (\ref{Vol integral}) and (\ref{rhat1}) implies that 
\beqnn
\{\tau_1>t\}=\Big\{\tau_1>t, \ \int_0^t\int_0^{V^{(y)}_{s-}}\int_{\overline{y}}^\infty\zeta N(ds,du,d\zeta)=0\Big\}\; a.s.
\eeqnn
If $\tau_1\leq t$,  we immediately have \beqnn
 \int_0^t\int_0^{V^{(y)}_{s-}}\int_{\overline{y}}^\infty\zeta N(ds,du,d\zeta)&\geq&
 \int_0^{\tau_1}\int_0^{V^{(y)}_{s-}}\int_{\overline{y}}^\infty\zeta N(ds,du,d\zeta)\\
 &=&\int_0^{\tau_1}\int_0^{V_{s-}}\int_{\overline{y}}^\infty\zeta N(ds,du,d\zeta)>0.
\eeqnn
Thus 
\beqlb\label{first large jump time}
\{\tau_1>t\}=\Big\{\int_0^t\int_0^{V^{(y)}_{s-}}\int_{\overline{y}}^\infty\zeta N(ds,du,d\zeta)=0\Big\}\; a.s.
\eeqlb

Recall that $1_{\{\zeta>\overline{y}\}}N(ds,du,d\zeta)$ is the restriction of $N(ds,du,d\zeta)$ to $(0,\infty)\times (0,\infty)\times (\overline{y},\infty)$, which is 
independent of $1_{\{\zeta\leq\overline{y}\}}N(ds,du,d\zeta)$. 
By (\ref{rhat1})  we have that  $1_{\{\zeta>\overline{y}\}}N(ds,du,d\zeta)$ is independent of $(V^{(y)}_t, t\geq 0)$.  Then
conditional on $(V^{(y)}_t, t\geq 0)$,  $\int_0^t \int_0^{V^{(y)}_{s-} } \int_{\overline y}^\infty {N}(ds,du,d\zeta)$  is a  time inhomogenous Poisson process 
with intensity function $\big(\int_{\overline y}^\infty\nu_\alpha(d\zeta)\big)V^{(y)}_{.}$. 
Note that  $\tau_1$ is the first jump time of $\sigma_{Z}\int_0^t \int_0^{V^{(y)}_{s-} } \int_{\overline y}^\infty {N}(ds,du,d\zeta)$, and $\Delta V_{\tau_1}$
is the first jump size of $\sigma_{Z}\int_0^t \int_0^{V^{(y)}_{s-} } \int_{\overline y}^\infty {N}(ds,du,d\zeta)$. Then we have 
\beqnn
\mathbb E\big[\tau_1\in dt,\ \Delta V_{\tau_1}\in d\zeta \,| V_{.}^{(y)}\big]
=\Big(\int_{\overline y}^\infty\nu_\alpha(dx) \Big)\,  \Big( V^{(y)}_{t}dt\Big) \, \Big(\frac{\alpha y^{\alpha}1_{\{\zeta>y\}}}{\zeta^{1+\alpha}}d\zeta\Big),
\eeqnn
which implies that $\Delta V_{\tau_1}$ is independent of $\tau_1$ and $V^{(y)}$. \finproof

{\noindent\bf Proof of Propositon \ref{Pro: decomposition}} 

Step 1. Recall that 
$
\tau_1=\inf\{t>0:\Delta V_t>y\}
$
and $T_1$ is the first jump time of the point process $\{J_t:t\geq0\}$ given by (\ref{mother jump intensity}).
By (\ref{first large jump time}),  we immediately get $\tau_1=T_1$ a.s.. Thus by Lemma \ref{large jump size}, we have 
that $V^{(y)}$ coincides with $V$ up to 
$T_1$ and $\Delta V_{T_1}$ is independent of $V^{(y)}$. Note that $V^{(y)}_{T_1}=V_{T_1-}$ and 
\begin{equation}\label{T1V}\begin{split}
{V}^{(y)}_t =V_{T_1-} &+ \int_{T_1}^t \widetilde{a}(\alpha, y) \big( \widetilde{b}(\alpha, y)  - {V}_s^{(y)}  \big) ds 
+ \sigma \int_{T_1}^t \int_0^{{V}_s^{(y)}} W(ds,du)\\
&+ \sigma_N \int_{T_1}^t \int_0^{{V}_{s-}^{(y)} } \int_0^{\overline{y}}\zeta \widetilde{N}
(ds,du,d\zeta), \quad t\geq T_1.
\end{split}\end{equation}
By taking $k=1$ in (\ref{cluster}), 
\begin{equation}\label{v1}\begin{split}
v_t^{(1)}=\Delta V_{T_1}  &-a \int_{T_1}^t v^{(1)}_s ds + \sigma \int_{T_1}^t 
\int_{{V}^{(y)}_s}^{{V}^{(y)}_s+ v^{(1)}_s} W(ds, du)\\
 &+ \sigma_N \int_{T_1}^t \int_{{V}^{(y)}_{s-}}^{{V}^{(y)}_{s-}+ v^{(1)}_{s-} } \int_{\mathbb{R}^+} 
\zeta \widetilde{N}(ds,du, d\zeta), \quad t\geq T_1. \end{split}\end{equation}
As mentioned above, $\Delta_{T_1}$ is independent of $V_{T_1-}$. By using the property of independent and stationary increments of $W$ and $N$, we have that $v^{(1)}$ and $V^{(y)}$ are independent of each other and $\{u^{(1)}_t:=v^{(1)}_{T_1+t},\ t\geq0\}$ is a CB process which has the same distribution as $u$ given by
(\ref{CB}); see e.g., \cite[Theorem 3.2, 3.3]{DawsonLi}). Now set 
 \begin{equation}\label{T1Vbar}\begin{split}
\bar{V}_t^{(1)} = V_{T_1-} 
&+ \int_{T_1}^t a \left( b  - \bar{V}_s^{(1)}  \right) ds +
 \sigma \int_{T_1}^t \int_0^{\bar{V}_s^{(1)}} W(ds, du)\\
&+ \sigma_N \int_{T_1}^t \int_0^{\bar{V}_{s-}^{(1)} }  \int_{\mathbb{R}^+}  \zeta \widetilde{N}
(ds,du, d\zeta),\quad t\geq  T_1.
\end{split}\end{equation}
It is easy to see $\bar{V}^{(1)}$  is of the same type  as $V$ but with initial value $V_{T_1-}$ and starting from time $T_1$. Define 
\beqnn
\bar{\tau}_1:=\inf\{t>T_1: \Delta \bar{V}_t^{(1)}>y\},
\eeqnn
which is the first jump time of $\bar{V}^{(1)}$ whose jump size larger than $y$.  Then a comparison of (\ref{T1V}) and (\ref{T1Vbar}) shows that $\bar{V}^{(1)}_t=V_t^{(y)}$ for $t\in[T_1,\bar\tau_1)$. Furthermore the similar proof of Lemma \ref {large jump size} shows that for any $t>0$, 
\beqnn
\{\bar{\tau}_1-T_1>t\}=\Big\{\int_{T_1}^{T_1+t}\int_0^{V^{(y)}_{s-}}\int_{\overline{y}}^\infty\zeta N(ds,du,d\zeta)=0\Big\}\; a.s.,
\eeqnn
which implies that $\bar\tau_1=T_2$ a.s. Thus $\Delta\bar{V}^{(1)}_{\bar\tau_1}=\Delta V_{T_2}$ and
$\Delta V_{T_2}$ is independent of $V^{(y)}$ and $\Delta V_{T_1}$.  Furthermore  $\bar{V}^{(1)}_t=V_t^{(y)}$ for $t\in[T_1,T_2)$.  We get that 
\beqlb\label{first interval}
V^{(y)}_t+v^{(1)}_t=\bar{V}_t^{(1)}+v^{(1)}_t=V_t, \ a.s.\quad t\in[T_1,T_2).
\eeqlb
The third equality follows from (\ref{T1Vbar}), (\ref{v1}) and (\ref{Vol integral}).

Step 2. By taking $k=2$ in (\ref{cluster}), 
\begin{equation}\label{v2}\begin{split}
v_t^{(2)}=\Delta V_{T_2}  &-a \int_{T_2}^t v^{(2)}_s ds + \sigma \int_{T_2}^t 
\int_{{V}^{(y)}_s+v^{(1)}_s}^{{V}^{(y)}_s+ v^{(1)}_s+v^{(2)}_s} W(ds, du)\\
 &+ \sigma_N \int_{T_2}^t \int_{{V}^{(y)}_{s-}+v^{(1)}_{s-}}^{{V}^{(y)}_{s-}+ v^{(1)}_{s-}+v^{(2)}_{s-}} \int_{\mathbb{R}^+} 
\zeta \widetilde{N}(ds,du, d\zeta), \quad t\geq T_2. \end{split}\end{equation}
Since $\Delta V_{T_2}$ is independent of $V^{(y)}_{T_2}$ and $\Delta V_{T_1}$, still by using the property of independent and stationary increments of $W$ and $N$, we have that $v^{(2)}$ are independent of $V^{(y)}$ and $v^{(1)}$,  and $\{u^{(2)}_t:=v^{(2)}_{T_2+t},\ t\geq0\}$ is also a CB process which has the same distribution as $u$. Now set 
 \begin{equation}\label{T1Vbar}\begin{split}
\bar{V}_t^{(2)} = V^{(y)}_{T_2} 
&+ \int_{T_2}^t a \left( b  - \bar{V}_s^{(2)}  \right) ds +
 \sigma \int_{T_2}^t \int_0^{\bar{V}_s^{(2)}} W(ds, du)\\
&+ \sigma_N \int_{T_2}^t \int_0^{\bar{V}_{s-}^{(2)} }  \int_{\mathbb{R}^+}  \zeta \widetilde{N}
(ds,du, d\zeta),\quad t\geq  T_2.
\end{split}\end{equation}
Define 
\beqnn
\bar{\tau}_2:=\inf\{t>T_2: \Delta \bar{V}_t^{(2)}>y\},
\eeqnn
As proved in Step 2 we have that $\bar\tau_2=T_3$ a.s. and $\bar{V}^{(2)}_t=V_t^{(y)}$ for $t\in[T_2,T_3)$.  Note that $V_{T_2-}=V^{(y)}_{T_2}+\Delta r^{(1)}_{T_2}$ by (\ref{first interval}). We get that 
\beqnn
V^{(y)}_t+v^{(1)}_t+v^{(2)}=\bar{V}_t^{(2)}+v^{(1)}_t+v^{(2)}_t=V_t, \ a.s.\quad t\in[T_2,T_3).
\eeqnn
Step 3. By induction,  it is not hard to prove that $V_t=V^{(y)}_t+\sum_{k=1}^nv^{(k)}_t$ holds for any $ t\in[T_n,T_{n+1})$  and $n\geq1$,  and the sequence of i.i.d processes is of the same distribution as $u$. Furthermore $\{u^{(n)}\}$ is independent of $V^{(y)}$. Then we have this proposition. \finproof

{\noindent\bf Proof of Propositon \ref{Pro: decomposition2}} \;(1) Note that $J_t^{(y)}\overset{d}{=}\int_0^t\int_0^{V_{s-}^{(y)}}\int_{D}M(ds,du, d\omega)$. Then 
\beqnn
\mathbb{E}[J_t^{(y)}]=\int_0^t\mathbb{E}[V_s^{(y)}]ds\int_{\bar y}^\infty\nu_{\alpha}(d\zeta).
\eeqnn
A simple computation shows (\ref{number cluster}). (2) By Proposition \ref{Pro: decomposition}, $u^{(n)}$ is a subcritical CB process without immigration, i.e. the branching mechanism is 
 \beqnn
 \Psi_{\alpha}(q)=a q+\frac{\sigma^2}{2}q^2-\frac{\sigma_N^\alpha}{\cos(\pi\alpha/2)}q^\alpha.
 \eeqnn
and the immigration rate $\Phi(q)=0$. Then $0$ is an absorbing point of $\theta_n$ and $\theta_n$ is the extinct time of CB process $u^{(n)}$. Since
$\int_1^{\infty}1/\Psi_{\alpha}(u)du<\infty$, the so-called Grey's condition is satisfied, it follows from Grey \cite[Theorem 1]{Grey74} that 
\beqnn
\mathbb{P}(\theta_n<\infty)=\int_0^\infty \mathbb{P}_x(\theta_n<\infty) \mathbb{P}(\Delta V_{T_n}\in dx)=1.
\eeqnn
Furthermore, still by \cite[Theorem 1]{Grey74}, we have that 
\beqlb\label{ODE2}
\mathbb{P}(\theta_n> t)=\mathbb{E}[1-e^{-{\Delta V_{T_n} q_t}}]=\alpha y^\alpha\int_{y}^{\infty}(1-e^{-xq_t})x^{-(1+\alpha)}dx,
\eeqlb
where $q_t$ is the minimal solution of the ODE
\beqnn
\frac{d}{dt}q_t=-\Psi_\alpha(q_t),\quad t>0,
\eeqnn
with $q_{0+}=\infty$. In this case, $0<q_t<\infty$ for $t\in(0,\infty)$. Then
\beqnn
\mathbb{E}[\theta_n]=\alpha y^\alpha\int_0^\infty\int_{y}^{\infty}(1-e^{-xq_s})x^{-(1+\alpha)}dxds,
\eeqnn
which gives (\ref{duration cluster}) by (\ref{ODE2}).\finproof

{\noindent\bf Proof of Propositon \ref{Pro: decomposition3}}
By (\ref{mother jump intensity}), we have
\beqnn
J_{nt}^{(y_n)}=\int_0^{nt}\int_0^{V^{(y_n)}_{s-}}\int_{\bar{y}_n}^\infty N(ds,du,d\zeta)
\eeqnn
where $\bar{y}_n=y_n/\sigma_N$. It follows from Proposition \ref{Pro: decomposition}-(2) that for any $\theta>0$,
\beqlb\label{Prop5.3 equality}
\mathbb{E}\Big[e^{-\theta J_{nt}^{(y_n)}}\Big]&=&\mathbb{E}\bigg[\exp\bigg\{
\Big(\int_{\bar{y}_n}^\infty \nu_\alpha(d\xi)\Big)\int_0^{nt}V_s^{(y_n)}ds\Big(e^{-\theta}-1\Big)\bigg\}\bigg]\nonumber\\
&=&
\mathbb{E}\bigg[\exp\bigg\{
\Big(n\int_{\bar{y}_n}^\infty \nu_\alpha(d\xi)\Big)\frac{1}{n}\int_0^{nt}V_s^{(y_n)}ds\Big(e^{-\theta}-1\Big)\bigg\}\bigg].
\eeqlb
Based on (\ref{rhat1}), for fixed $y_n$, $\{V_t^{(y_n)}:t\geq0\}$ is a CBI process. By \cite[Remark 5.3]{JMS2017}, for $\theta>0$, 
\beqlb\label{LS of truncated processes}
\mathbb{E}\bigg[e^{-\frac{\theta}{n}\int_0^{nt} V_s^{(y_n)}ds}\bigg]=\exp\bigg\{-v_n(\theta,nt)V_0-ab\int_0^{nt}v_n(\theta,s)ds\bigg\}
\eeqlb
where $v_n(\theta,t)$ is the unique solution of 
\beqlb\label{ODE of n}
\frac{dv_n(\theta,t)}{dt}=\frac{\theta}{n}-\Psi_n(v_n(\theta,t)),
\eeqlb
with $v_n(\theta,0)=0$, and 
\beqnn
\Psi_n(q)=\Big(a+\sigma_N^\alpha\int_{y_n}^\infty\xi\nu_\alpha(d\xi)\Big)q+\frac{\sigma^2}{2}q^2+\sigma_N^\alpha\int_0^{y_n}
(e^{-q\xi}-1+q\xi)\nu_\alpha(d\xi).
\eeqnn
Then we have $-\Psi_n(v_n(\theta,t))\leq\frac{d v_n(\theta,t)}{dt}\leq \frac{\theta}{n}-av_n(\theta,t)$, which implies that $0\leq v_n(\theta,t)\leq \frac{\theta}{
an}(1-e^{-at})$. By (\ref{ODE of n}),
\beqlb\label{ODE of n 2}
nv_n(\theta,nt)=\frac{\theta}{a_n}(1-e^{-na_nt})-\int_0^{nt}e^{-a_n(nt-s)}n\hat{\Psi}_n(v_n(\theta,s))ds,
\eeqlb
where 
\beqnn
a_n=a+\sigma_N^\alpha\int_{y_n}^\infty\xi\nu_\alpha(d\xi),\quad \hat{\Psi}_n(q)=\frac{\sigma^2}{2}q^2+\sigma_N^\alpha\int_0^{y_n}
(e^{-q\xi}-1+q\xi)\nu_\alpha(d\xi).
\eeqnn
Note that $a_n\rightarrow a$, and for all $t\geq 0$ and $n\geq1$,
\beqnn
0\leq nv_n(\theta,t)\leq \frac{\theta}{a}, \quad n\hat{\Psi}_n(v_n(\theta,t))\leq \frac{\sigma^2\theta^2}{2a^2n}-\frac{\sigma_N^\alpha\theta^\alpha}{\cos(\pi\alpha/2)a^\alpha n^{\alpha-1}}.
\eeqnn
By (\ref{ODE of n 2}), we have $nv_n(\theta,nt)\rightarrow\frac{\theta}{a}$ and then
\beqnn
\int_0^{nt}v_n(\theta,s)ds=\int_0^tnv_n(\theta,ns)ds\rightarrow\frac{\theta t}{a}.
\eeqnn
Thus by (\ref{LS of truncated processes}), we have for any $t\geq0$,
\beqnn
\frac{\int_0^{nt}V_s^{(y_n)}ds}{n}\overset{p}{\rightarrow} bt.
\eeqnn
Recall that $y_n\sim cn^{1/\alpha}$. Then $n\int_{\bar{y}_n}^\infty \nu_\alpha(d\xi)\rightarrow-
\frac{\sigma_N^\alpha}{\alpha\cos(\pi\alpha/2)\Gamma(-\alpha)c^\alpha}$. By (\ref{Prop5.3 equality}),
\beqnn
\mathbb{E}\Big[e^{-\theta J_{nt}^{(y_n)}}\Big]\rightarrow \exp\bigg\{-\frac{\sigma_N^\alpha bt}{\alpha\cos(\pi\alpha/2)\Gamma(-\alpha) c^\alpha}(e^{-\theta}-1)\bigg\}.
\eeqnn
We are done.

\bibliographystyle{spbasic}

\begin{thebibliography}{99}

\bibitem{AP2017} Abi Jaber, E., Larsson, M.  and Pulido, S. (2017): Affine Volterra processes. {\it Working Paper}

%\bibitem{AJ09} A\"it-Sahalia, Y. and Jacod, J. :Estimating the Degree of Activity of Jumps in High Frequency Data. \textit{Annals of Statistics}, 37, 2202-2244 (2009).
%
%\bibitem{AJ15} A\"it-Sahalia, Y., Cacho-Diaz, J. and Laeven, R. J.: 
%Modeling financial contagion using mutually exciting jump processes. \textit{Journal of Financial Economics}.
%117(3), 585-606  (2015).
%
%\bibitem{ALM}  Albeverio, S.,  Lytvynov, E. and  Mahnig, A.: A model of the term structure of interest rates based on L\'evy fields. {\it Stochastic Processes and their Applications}, 114, 251-263 (2004).
%
%

\bibitem{AP2018} Avellaneda, M. and Papanicolaou, A.  (2018): Statistics of VIX Futures and Applications to Trading Volatility Exchange-Traded Products. Journal of Investment Strategies,  7(2), 1-33. 

\bibitem{AR2001} Asmussen, S. and Rosinski, J.: Approximations of small jumps of L\'evy processes with a view towards simulation, {\it Journal of Applied Probability}, 38, 482-493  (2001).


%\bibitem{Axtell01} Axtell, R. L.: Zipf distribution of US firm sizes. {\it Science}, 293(5536), 1818-1820 (2001).

%\bibitem{BB14} Baldeaux, J., and  Badran, A. (2014). Consistent modelling of VIX and equity derivatives using a 3/2 plus jumps model. Applied Mathematical Finance, 21(4), 299-312.

\bibitem{BNS2001} Barndorff-Nielsen, E. and Shephard, N. (2001): Non-Gaussian Ornstein-Uhlenbeck-based models and some of their uses in financial economics, {\it Journal of Royal Statistical Society, Series B}, 63(2), 167-241.

\bibitem{Bates1996} Bates, D. (1996): Jump and stochastic volatility: exchange rate processes implicit in Deutsche market options, {\it Review of Financial Studies}, 9, 69-107. 


\bibitem{BFG2016}  Bayer, C.; Friz, P.,  and Gatheral, J. (2016): Pricing under rough volatility. \textit{Quantitative Finance},
16(6), 887-904.

\bibitem{BGT87} Bingham, N. H., Goldie C. M. and Teugels, J. L. (1987): Regular Variation. Cambridge: Cambridge Univ Press.

\bibitem{BT75}Bingham, N. H. and Teugels, J. L. (1975): Duality for regularly varying functions. {\it Quart J Math Oxford}, 26, 333-353.





%
%
%\bibitem{BS01}
%Barndorff-Nielsen, O.E. and Shephard, N.: 
%Modelling by L\'evy processes for financial econometrics, in {\it L\'evy 
%Processes: Theory and Applications}, eds. Barndorff-Nielsen, Mikosch and Resnick, Birkh\"auser Verlag GmbH, 283-318 (2001).
%
%
%
%\bibitem{BD86} Brown, S. J. and Dybvig, P. H.: 
%The empirical implications of the Cox, Ingersoll, Ross theory of the term structure of interest rates. \textit{Journal of Finance}, 41, 617-630  (1986).
%

%\bibitem{CGSS2017} Callegaro, G., Ga\"igi, M., Scotti, S. and Sgarra, C. (2017). Optimal investment in markets with over and under-reaction to information. \textit{Mathematics and Financial Economics}, 11 (3), 299-322.

%\bibitem{CM1999} Carr, P., and Madan, D. (1999). Option valuation using the fast Fourier transform. \textit{Journal of computational finance}, 2(4), 61-73.

\bibitem{CL2007} Carr, P. and Lee, R. (2007): Realized volatility and variance: options via swaps, {\it RISK}, 20, 76-83.






\bibitem{CIR85} Cox, J., Ingersoll, J. and Ross, S.: A theory of the term structure of interest rate. \textit{Econometrica}, {53}, 385-408  (1985).


\bibitem{DaFGra11} Da Fonseca, J. and Grasselli, M. (2011): Riding on the smiles. {\it Quantitative Finance} 11.11 (2011): 1609-1632.

%\bibitem{DZ2011} Dassios, A. and Zhao, H.: A dynamic contagion process, {\it Advances in Applied Probability}, 43(3), 814-846  (2011).
%
%\bibitem{DL06} Dawson, D.A. and Li, Z.: Skew convolution semigroups and affine Markov processes. \textit{Annals of Probability},{34}, 1103-1142  (2006).
% 

\bibitem{DawsonLi} Dawson, A and Li, Z.: Stochastic equations, flows and measure-valued processes. \textit{Annals of Probability}, 40(2), 813-857  (2012).

%\bibitem{Drimus2011} Drimus G. (2012): Option on realized variance by transform methods: a non-affine stochastic volatility model, {\it Quantitative finance}, 12(11), 1679-1694.

\bibitem{DPS2000} Duffie, D., Pan, J. and Singleton, K. (2000): Transform analysis and asset pricing for affine jump-diffusions. {\it Econometrica}, 68(6), 1343-1376.


%
%\bibitem{DG2001}Duffie, D. and G\^arleanu, N.: Risk and valuation of collateralized debt obligations. {\it Financial Analysts Journal}, 57(1), 41-59  (2001).
%
\bibitem{DFS2003}
Duffie, D., Filipovi\'{c}, D. and Schachermayer, W.: Affine processes and applications in finance, {\it Annals of Applied Probability}, 13(3), 984-1053  (2003).

\bibitem{DuqLab}
Duquesne, T. and Labb{\'e}, C. (2014):  On the {E}ve property for {CSBP},
  {\it Electron. J. Probab}. \textbf{19}, no. 6, 31 pages.




%
%
%
%
%\bibitem{DFM14}
%Duhalde, X.,  Foucart, C. and Ma, C.: {On the hitting times of continuous-state branching processes with immigration}, {\it Stochastic Processes and their Applications}, 124(12), 4182-4201  (2014). 
%
%
%
%\bibitem{EberleinRaible1999} Eberlein, E.  and  Raible, S.: Term structure models driven by general L\'evy processes. {\it Mathematical Finance},  9, 31-53  (1999).
%
%\bibitem{KM90}
%El Karoui, N. and M\'{e}l\'{e}ard, S.: Martingale measures and
%stochastic calculus. \textit{Probability Theory and Related Fields}, 84, 83--101 (1990).
%

\bibitem{ER2016} El Euch, O. and Rosenbaum M. (2016): The characteristic function of rough Heston model, preprint to appear in {\it Mathematical Finance}.
\bibitem{EFR2016}  El Euch, O., Fukasawa,  M. and Rosenbaum M. (2016) The microstructural foundations of leverage effect and rough volatility, preprint to appear in {\it Finance and Stochastics}.



\bibitem{EGG2010} Errais, E., Giesecke, K., Goldberg, L.  (2010): Affine point processes and portfolio credit risk. {\it SIAM Journal on Financial
Mathematics}, 1, 642-665.
%
%\bibitem{EK86}
%Ethier, S.N. and Kurtz, T.G.: \textit{Markov Processes:
%Characterization and Convergence}, John Wiley and Sons, New York  (1986).
%
%
%
%

%\bibitem{FKL2006} Fasen V., Klüppelberg C., Lindner A. (2006) Extremal behavior of stochastic volatility models. In: Shiryaev A.N., Grossinho M.R., Oliveira P.E., Esquível M.L. (eds) Stochastic Finance. Springer

\bibitem{F01} Filipovi\'{c}, D.  (2001): A general characterization of one factor affine term structure models, \textit{Finance and  Stochastics}, 5(3), 389-412.
%
%\bibitem{F02} Filipovi\'{c}, D.: {\it Term Structure Models}, Springer, New York  (2009). 
%
%
%\bibitem{FilipovicTappeTeichmann2010} Filipovi\'c, D.,  Tappe, S. and Teichmann, J.: Term structure model driven by Wiener process
%and Poisson measures: existence and positivity, {\it SIAM Journal on Financial Mathematics}, 1, 523-554  (2010).
%
% 
%
\bibitem{FL10} Fu, Z. and Li, Z.   (2010): Stochastic equations of non-negative processes with jumps. \textit{Stochastic Processes and their Applications}, {120}, 306-330.
%

\bibitem{GJR2014} Gatheral, J.,   Jaisson, T. and Rosenbaum, M. (2014): Volatility is rough. Preprint to appear  in {\it Quantitative Finance}, arXiv:1410.3394.


\bibitem{GM2013} Goard, J., and Mazur, M. (2013). Stochastic volatility models and the pricing of VIX options. Mathematical Finance, 23(3), 439-458.
%\bibitem{GR93}
%Gibbons, M. R., and Ramaswamy, K.: 
%A test of the Cox, Ingersoll and Ross model of the term structure. \textit{Review of Financial Studies}, 6(3), 619-658  (1993).
%
%\bibitem{Gopikrishnan99} Gopikrishnan, P., Plerou, V., Amaral, L. A. N., Meyer, M., and Stanley, H. E. (1999): Scaling of the distribution of fluctuations of financial market indices. {\it Physical Review E}, 60, 5305-5316.


%\bibitem{GT2008} Grasselli, M., and Tebaldi, C. (2008). Solvable affine term structure models. \textit{Mathematical Finance}, 18(1), 135-153.

\bibitem{G2016} Grasselli, M. (2016). The 4/2 stochastic volatility model: a unified approach for the Heston and the 3/2 model. \textit{Mathematical Finance}.

\bibitem{Grey74} Grey, D.R. (1974): Asymptotic behavior of continuous time continuous state-space branching processes. \textit{J. Appl. Prob.} \textbf{11}, 669-677.

\bibitem{Guli2010} Gulisashvili, A. (2010): Asmptotic formulas with error estimates for call pricing functions and the implied volatility at extreme strikes. \textit{Siam J. Financial Math.} \textbf{1},609-641.

\bibitem{Hawkes} Hawkes, A. G.   (1971): Spectra of Some Self-Exciting and Mutually Exciting Point Processes. \textit{Biometrika}, 58, 83-90.
%
%\bibitem{HL16} He, X. and  Li, Z.  (2016): Distributions of jumps in a continuous-state branching process with immigration. {\it Journal of Applied Probability}, 53(4), 1166-1177.
%

\bibitem{Hes1993} Heston, S. (1993): A closed form solution for options with stochastic volatility with applications to bond
and currency options,  \textit{Review of Financial Studies} 6, 2, 327-344.

\bibitem{HL07}Hult, H. and Lindskog, F. (2007): Extremal behavior of stochastic integrals driven by regularly L\'{e}vy processes.\textit{Ann. Probab.} \textbf{35}, 309-339.



\bibitem{JR15} Jaisson, T., and Rosenbaum, M.   (2015): Limit theorems for nearly unstable Hawkes processes. \textit{Annals of Applied Probability}, 25(2), 600-631.


%\bibitem{JKWW2011} Janek, A., Kluge, T., Weron, R., and Wystup, U. (2011): FX smile in the Heston model. \textit{Statistical tools for finance and insurance}, 133-162.

\bibitem{JMS2017} Jiao, Y., Ma, C. and Scotti S. (2017): Alpha-CIR model with branching processes in sovereign interest rate modeling,
\textit{Finance and Stochastics}, 21(3), 789-813.

%\bibitem{JMSS2017} Jiao, Y., Ma, C., Scotti S. and Sgarra, C. (2017): A Branching Process Approach to Power Markets, submitted REFERENCE.


%
\bibitem{KMK2010}
Kallsen, J. and Muhle-Karbe, J.: Exponentially affine martingales, affine measure changes and exponential moments of affine processes, {\it Stochastic Processes and their Applications}, 120, 163-181  (2010).
%

\bibitem{KMV2011} Kallsen, J., Muhle-Karbe, J. and Vo$\beta$, M. (2011): Pricing option on variance in affine stochastic volatility models,
\textit{Mathematical Finance}, 21, 627-641.

%\bibitem{KaW71} Kawazu, K. and Watanabe, S.: Branching processes with immigration and related limit theorems. \textit{Theory of Probability and its Applications}, {16}, 36-54  (1971).


\bibitem{KR11} Keller-Ressel, M. (2011): Moment explosions and long-term behavior of affine stochastic volatility models. {\it Mathematical Finance}, 21, 73-98.

%
%\bibitem{KRM12}Keller-Ressel M. and Mijatovi\'c, A.: On the limit distributions of continuous-state branching processes with immigration, {\it Stochastic Processes
%and their Applications}, 122(6), 2329-2345  (2012).
%
%\bibitem{KRS08}Keller-Ressel, M. and Steiner, T.: Yield curve shapes and the asymptotic short rate distribution in affine one-factor models. \textit{Finance and Stochastics}, 12(2), 149-172  (2008). 
%
%\bibitem{Ken} Kennedy, D.:  The term structure of interest rates as a Gaussian random field. {\it Mathematical Finance}, 4, 247-258  (1994).
%
%

\bibitem{L04} Lee, R. W.: The moment formula for implied volatility at extreme strikes. {\it Mathematical Finance}, 14(3), 469-480  (2004).

%
%
\bibitem{Li11} Li, Z.: \textit{Measure-Valued Branching Markov Processes}, Springer, Berlin  (2011).
%
%
\bibitem{LM13}Li, Z. and Ma, C.:  Asymptotic properties of estimators in a stable Cox-Ingersoll-Ross model.   {\it Stochastic Processes and their Applications},  {125}(8), 3196-3233  (2015).
%
%



\bibitem{Liu99} Liu, Y., Gopikrishnan, P., Cizeau, P., Meyer, M., Peng, C. K. and Stanley H.E. (1999): Statistical properties of the volatility of price fluctuations. {\it Physical Review E}, 60, 1390-1400.

\bibitem{Lon10} Long, H. (2010): Parameter estimation for a class of stochastic differential equations driven by small stable noises from discrete observations. \textit{Acta Math. Sci. English Ed.}{30B}, 645-663.


\bibitem{NV03}  Nicolato, E. and Venardos, E.  (2003):  Option pricing in stochastic volatility models of the Ornstein-Uhlenbeck type, {\it Mathematical Finance}, 13 , 445-466.

\bibitem{NPS2017}Nicolato, E., Pisani, C. and Sloth, D. (2017): The impact of jump distributions on the implied volatility of variance, {\it SIAM Journal on Financial Mathematics}, 8, 28-53.

%\bibitem{RPL14} Rambaldi, M., Pennesi, P., and Lillo, F.: Modeling FX market activity around macroeconomic news: a Hawkes process approach. {\it Physical  Review E}, 91, 012819  (2015).

\bibitem{RW85} Rosinski, J. and Woyczynski, W.A. (1985): Moment
inequalities for real and vector p-stable stochastic integrals. In:
\textit{Probability in Banach Spaces V}. Lecture Notes in Math.
{1153}, 369-386, Springer, Berlin.



\bibitem{Sepp2008} Sepp, A. (2008): Pricing options on realized variance in the Heston model with jumps in returns and volatility, {\it Journal of Computational Finance}, 11, 33-70.

%\bibitem{Sepp2008-2} Sepp, A. (2008): VIX option pricing in a jump diffusion model, {\it Risk}, 84-89, (April 2008).


%
%
\end{thebibliography}

\end{document}